\documentclass[pdflatex,sn-nature]{sn-jnl}

\usepackage{graphicx}
\usepackage{overpic}
\usepackage{subcaption}
\usepackage{tikz}
\usepackage{amsmath,amssymb,amsfonts}
\usepackage{bm}
\usepackage{booktabs}
\usepackage{float}
\usepackage{placeins}
\usepackage[title]{appendix}
\usepackage[mathlines]{lineno}
\DeclareCaptionLabelFormat{panel}{#2.}
\makeatletter
\def\email#1{\global\advance\emailcnt by 1\relax%
\if@corauemail%
   \g@addto@macro\corrauthemail{%
   \setcounter{footnote}{0}%
   \textcolor{blue}{#1}\ %
   }%
\else%
   \g@addto@macro\authemail{%
   \setcounter{footnote}{0}%
   \textcolor{blue}{#1}\ %
   }%
\fi}
\makeatother

\begin{document}

\title[Foundation model for traffic flow dynamics]{A learning framework toward foundation models for road-link traffic dynamics}

\author[1]{\fnm{Xiaoyu} \sur{Ma}}
\author*[1,2]{\fnm{Sean} \sur{Qian}}\email{seanqian@cmu.edu}

\affil[1]{\orgdiv{Department of Civil and Environmental Engineering}, \orgname{Carnegie Mellon University}, \orgaddress{\city{Pittsburgh}, \state{PA}, \postcode{15213}, \country{USA}}}
\affil[2]{\orgdiv{Heinz College of Information Systems and Public Policy}, \orgname{Carnegie Mellon University}, \orgaddress{\city{Pittsburgh}, \state{PA}, \postcode{15213}, \country{USA}}}

\abstract{Traffic flow models are central to transportation operations and planning, but conventional macroscopic and mesoscopic formulations often rely on limited variables, rigid functional forms, and location-specific calibration. These constraints limit their ability to represent the heterogeneous interactions among traffic states, road geometry, vehicle composition, land use, weather, and other contextual conditions. Here, we propose a Transformer-based foundation-model framework for learning link-level traffic flow dynamics from large-scale, multi-source data, and demonstrate it as a proof of concept using comprehensive regional datasets. The framework combines self-supervised pre-training, which learns latent representations of road links and their operating conditions, with supervised fine-tuning for traffic regime learning as an example downstream task. The trained model achieves higher prediction accuracy on held-out observations and exhibits stronger zero-shot transferability to unseen regions, compared with conventional physics-based and machine-learning baselines. Analyses of the learned latent space show that fine-tuning organizes traffic states into physically meaningful regimes, recovers structures resembling the fundamental diagram, and assigns larger representational updates to congested states. Feature evaluation and attention visualization further validate the model's advantages by integrating dynamic flow variables with geometric and environmental factors such as precipitation, sinuosity, grade, and truck proportion. By replacing non-differentiable hand-coded traffic models with a differentiable, transferable neural representation, this work serves as a proof-of-concept envisioning broader impacts on providing a computational foundation for scalable traffic estimation, simulation, and future optimization-based transportation planning, operations, and control.}


\maketitle

\section{Introduction}\label{sec:introduction}

Modeling, learning and prediction of traffic flow dynamics of road segments (\textit{road links} will be used interchangeably) is a cornerstone of modern transportation operation and planning systems. The essential element is to study how vehicles of various attributes move through road networks, examine the relationships between speed, density, and flow to understand and predict traffic breakdown, travel times across segments, and the overall behavior of traffic under varying conditions. It is well known that subtle changes in vehicle car-following or lane-change can induce traffic breakdown, unstable traffic states, and tipping-point behavior, all of which are central to traffic resilience and underscore the importance of learning traffic flow regimes underlying complex nonlinear dynamics \cite{zeng2020multiple, olmos2018macroscopic}.

\subsection{Motivation and Research Gaps}

With decades of research, the basic mechanism of microscopic vehicle car-following or lane-change behaviors is understood. However, predicting and managing traffic flow at the system level or mesoscopic level remains challenging, due to the nature of sensitivity, randomness and heterogeneity in traffic flow. For instance, fundamental diagrams \cite{greenshields1935study, lighthill1955kinematic, daganzo1997fundamentals} of traffic flow rate, space mean speed and density are widely recognized, but largely exhibit random and volatile nature. There is no mathematical formulation that well defines fundamental diagrams for real-world traffic. This paper aims to establish a new concept of foundation models for traffic flow dynamics under various conditions of built and natural environment that can replicate the real-world traffic propagation at the mesoscopic or macroscopic level (namely, traffic of analysis interest are aggregated spatially on road segments, and temporally in a few seconds or minutes).

Traditional macroscopic or mesoscopic traffic flow models \cite{greenshields1935study, lighthill1955kinematic, richards1956shock, haight1965mathematical, friesz1993variational, newell1993simplified, jin2021link, daganzo1994cell} and data-driven traffic state estimation or prediction models \cite{sun2014data, mori2015review, oh2018short, bramich2022fitting, wang2005real, seo2017traffic, xing2022traffic, van2002freeway, ma2015long, yu2017spatio, cui2019traffic, huang2020physics, di2023physics, tedjopurnomo2020survey, yin2021deep, qu2025transformer}, either spatial-temporal modeling or time-series prediction, have focused predominantly on limited static and dynamic features (such as flow, density, and flow/jam capacity) to model or estimate traffic states of road segments. Eq.~(\ref{abstract func - traditional}) shows an abstract functional form of the traditional macroscopic or mesoscopic traffic flow models, where \(\hat{y}(t)\) denotes the traffic state of a road link at time \(t\). \(\pmb{\alpha}\) represents the set of time-invariant link characteristics such as capacity and free-flow speed. \(\pmb{\beta}(t)\) represents the set of dynamic features such as flow and density at time \(t\) . Specifically, for example, in the delay-function link model \cite{friesz1993variational} and the Greenshields model \cite{greenshields1935study}, \(\Gamma\) is a linear or non-linear scalar function. In dynamic link-queuing models, such as the link transmission model \cite{newell1993simplified}, cell transmission model \cite{daganzo1994cell}, or link queue model \cite{jin2021link}, \(\Gamma\) is a system of partial differential equations or ordinary differential equations that determines the cumulative flow curves for calculation of link travel time or other traffic state.

\begin{equation}\label{abstract func - traditional}
    \hat{y}(t) \leftarrow \Gamma (\pmb{\alpha}, \pmb{\beta}(t))
\end{equation}

While these variables considered in traditional models are impeccably central to traffic kinematics, the physical traffic propagation traversing a road network is far more complex. In practice, the resulting link traffic states are fundamentally constrained and influenced by a multitude of external factors, including road geometric design (e.g., grades, sinuosity, land use, lane width), multi-class vehicle composition (e.g., truck proportion, make/year/model), traffic dynamics (e.g., flow, density), and environmental conditions (e.g., precipitation, visibility, holidays). Traditional models, often constrained by rigid functional forms or extremely limited input dimensionality, struggle to fuse these heterogeneous, multi-source data streams. Consequently, they frequently fail to capture the highly non-linear, asymmetric interactions between static infrastructure, dynamic traffic states, and contextual or environmental conditions, leading to inability to effectively and accurately model traffic flow dynamics, particularly degraded performance under anomalous conditions. In addition, traditional models are typically calibrated using data from extremely limited locations and time, and thus can hardly be transferred to other locations. 

One such series of traditional traffic flow models is summarized in the Highway Capacity Manual (HCM) \cite{trb2022hcm}. HCM used limited data to calibrate a speed-flow fundamental diagram, which is further parameterized by various conditions, such as truck composition, grades, highway/freeway configurations, etc. These fundamental diagrams are practically a reasonable approximation of flow characteristics, including the flow capacity, but are not sufficiently accurate for real-time traffic flow operation or high-fidelity infrastructure planning. Traditional macroscopic or mesoscopic dynamic traffic flow models, built on top of those approximate fundamental diagrams, are established to encapsulate time-varying flow dynamics. One outstanding issue is that they represent a general governing rule of how traffic flow, analogous to kinetic waves and with extremely limiting assumptions (e.g., homogeneous vehicle composition, vehicle-infrastructure interaction is largely captured by setting the flow capacity), but those dynamics models in their current standing are not adequately accurate to replicate real-world time-varying flow dynamics. 

Furthermore, classical macroscopic/mesoscopic dynamic traffic flow models that reflect certain dynamic queuing behaviors are computationally expensive and fundamentally non-differentiable. This non-differentiability creates a mathematical bottleneck for traffic management and optimization problems in large-scale networks, such as network flow routing, dynamic pricing, adaptive signal control, just to name a few, which rely on these models to simulate underlying constraints governing flow propagation but cannot easily apply efficient gradient-based solvers to them.

Recent advances in deep learning, particularly the emergence of foundation models and attention-based architectures \cite{vaswani2017attention}, offer a potential paradigm shift for the modeling of traffic flow of road links. Unlike conventional models trained from scratch on narrow, task-specific datasets, foundation models leverage large amounts of unlabeled data to learn highly generalized, reusable representations. By mapping complex inputs into a high-dimensional latent space, these models can uncover deep underlying correlations that dictate generic system behavior, which, in this case, lies in how traffic flow with different flow composition interacts with the built and natural environment under varying conditions. Leveraging these advances, representation learning and generative models have been widely utilized across various problems in the transportation field \cite{xu2025predicting, chen2025temporal, yuan2025learning, li2025activity, han2021multi, chen2021robust, wang2019learning}, demonstrating their strong potential.

With the support from multi-source data, foundation models and representation learning present an unprecedented opportunity: rather than explicitly hard-coding the physics models or manually tuning influential factors (e.g., those from the HCM), we can train a neural architecture to autonomously learn the foundational physics of traffic dynamics directly from a comprehensive, multi-source dataset, meanwhile enhancing potential transferability across regions. Crucially, by replacing rigid, non-differentiable analytical models with a pure neural architecture, the entire traffic simulation and estimation process becomes fully differentiable, while remaining consistent with the true underlying physics of flow dynamics. This end-to-end differentiability bridges the gap between flow estimation and control, facilitating the development of highly accurate, scalable, and gradient-friendly downstream optimization applications, particularly for large-scale networks. Furthermore, the self-attention mechanisms inherent in these models provide potential for physical interpretability, allowing researchers to peer into the ``black box" and understand exactly how different factors dynamically route information to estimate macroscopic traffic pattern.

This research focuses on learning latent representations and traffic regimes at the road-link level. The closest work to this study is a prior study \cite{choudhury2024scalable} which explored the learning of link-level traffic congestion functions using traditional feed-forward neural networks, incorporating a set of static and dynamic features. While that work leverages neural network architectures for traffic modeling, its capacity is limited to learning the superficial relationships between travel time/speed and the considered factors. It fails to learn the underlying physics and the foundational mechanisms of traffic flow due to the limited set of traditional factors (without considering a wide range of natural and build environmental factors and network spatial-temporal interactions) and a classical, less sophisticated neural network structure, which can lead to degraded performance especially when it is transferred to different regions or varying scenarios. 

\subsection{A Closer Look at the Proposed Work}

In this study, we consider a comprehensive range of factors that play an important role in traffic flow dynamics, including static physical characteristics of the road infrastructure, dynamic traffic flow characteristics, and contextual conditions such as weather (details are provided later). More importantly, we leverage recent advances in latent space representation learning, foundation models, and attention-based architectures. Such a comprehensive feature design, combined with advanced modeling techniques, is essential for capturing the underlying physical mechanisms of traffic and environment interactions and improving the model's transferability. 

The factors considered include the following three types. For detailed information about the definitions, data, and relevant calculations regarding these factors, please refer to the Appendix.

\textbf{\textit{Static features}} (\(\tilde{\pmb{\alpha}}\), time-invariant): Representing the physical characteristics of the road segments, including length, free-flow speed, number of lanes, number of in/out links, orientation, grade, sinuosity, maximum and mean curvature, and the surrounding land use pattern.

\textbf{\textit{Dynamic features}} (\(\tilde{\pmb{\beta}}\), time-varying): Representing the dynamic traffic states, including the time-varying traffic flow, density, speed, and truck proportion. It may also include the dynamic traffic state information of neighboring links. 

\textbf{\textit{Contextual features}} (\(\pmb{\gamma}\), time-varying): Representing the situational and environmental context, including time of day, day of week, holiday status, temperature, weather type, precipitation, and visibility.

We choose those listed features as they become ubiquitously and inexpensively available. Since the methodology for training new foundation models is generic, additional features potentially influencing traffic flow could be further added in the future as we continue to improve the models. 

Eq.~(\ref{abstract func - ours}) provides an abstract functional form of the proposed model. \(\hat{y}_a(t)\) denotes the traffic state (i.e., travel time, speed, density, or flow) of link \(a\) at time \(t\).  Compared with traditional macroscopic (or mesoscopic) traffic flow models and existing relevant studies, the static features (\(\tilde{\pmb{\alpha}}, \pmb{\alpha} \subset \tilde{\pmb{\alpha}}\)) covers more geometric features of the road itself (such as number of in/out links, orientation, grade, sinuosity, curvature, and surrounding land use). The dynamic features (\(\tilde{\pmb{\beta}}, \pmb{\beta} \subset \tilde{\pmb{\beta}}\)) also incorporate the temporal interrelationships of dynamic traffic states (i.e., dynamics back to any past time stamp \(t_0\):  \(\{t'|t' \in [t_0,t]\}\)), vehicle-to-vehicle interrelationships (vehicle heterogeneity), and dynamic traffic states of neighboring links (denote \(N_a\) as the set of neighboring links). In addition, we also innovatively consider a set of time-varying contextual features (\(\pmb{\gamma}\)) that reflect natural or social environments. All these considerations play a critical role in influencing road traffic but have been largely overlooked in existing studies.

\begin{align}\label{abstract func - ours}
    & \hat{y}_a(t) \leftarrow \pmb{\Gamma}_{\theta_2}\Big(\phi_{\theta_1} \big(\tilde{\pmb{\alpha}}_a, \tilde{\pmb{\beta}}_a(t'|t' \in [t_0,t]), \tilde{\pmb{\beta}}_{a'\in N_a}(t'|t' \in [t_0,t]), \pmb{\gamma}(t)\big)\Big) \\
    & \phi_{\theta_1}: \mathbb{R}^{d} \rightarrow \mathbb{R}^{D}, \quad D > d \nonumber
\end{align}

\(\phi_{\theta_1}\) and \(\pmb{\Gamma}_{\theta_2}\) are neural network models parameterized by \(\theta_1\) and \(\theta_2\), respectively. They are trained using real-world data (including those input features and observed link travel times). Specifically, the proposed learning framework follows a standard foundation model pipeline consisting of two distinct phases: a self-supervised pre-training phase (\(\phi\)) designed to learn fundamental representations of road links under diverse circumstances, followed by a supervised fine-tuning phase (\(\pmb{\Gamma}\)) for link-level traffic flow regime learning. The output of the pre-training phase is a dense, high-dimensional embedding for each data sample, which can act as a reusable latent representation to be further fine-tuned for various downstream applications. In this research, we focus solely on one downstream application as an example: analyzing the traffic flow regime via link-level traffic state estimation.

Note that the proposed framework is a purely data-driven approach that does not impose any predefined functional forms or explicitly incorporate traffic flow physics. In the current implementation, traffic states are represented at an aggregated road-link level, where the spatial definition of a link is typically affected by the available real-world measurements (e.g., sensor locations and Traffic Message Channel (TMC) segment definitions). However, if high-resolution data become available, such as second-by-second GPS observations or vehicle trajectories, the proposed framework could be extended to, by much shorter 'links', represent traffic dynamics at much finer spatial and temporal resolutions. As the spatial and temporal discretization approaches infinitesimal scales, the framework could potentially approximate continuous traffic dynamics and capture phenomena such as queue formation, spillback, and shockwave propagation. This would achieve a level of traffic dynamics representation comparable to that of traditional partial differential equation (PDE)-based traffic flow models, without explicitly prescribing governing physical equations. Nevertheless, the appropriate level of spatial and temporal resolution should ultimately be determined by practical application needs, as many real-world transportation applications do not necessarily require continuous representations of traffic dynamics.

\begin{figure}[t]
\centering
\includegraphics[width=\textwidth]{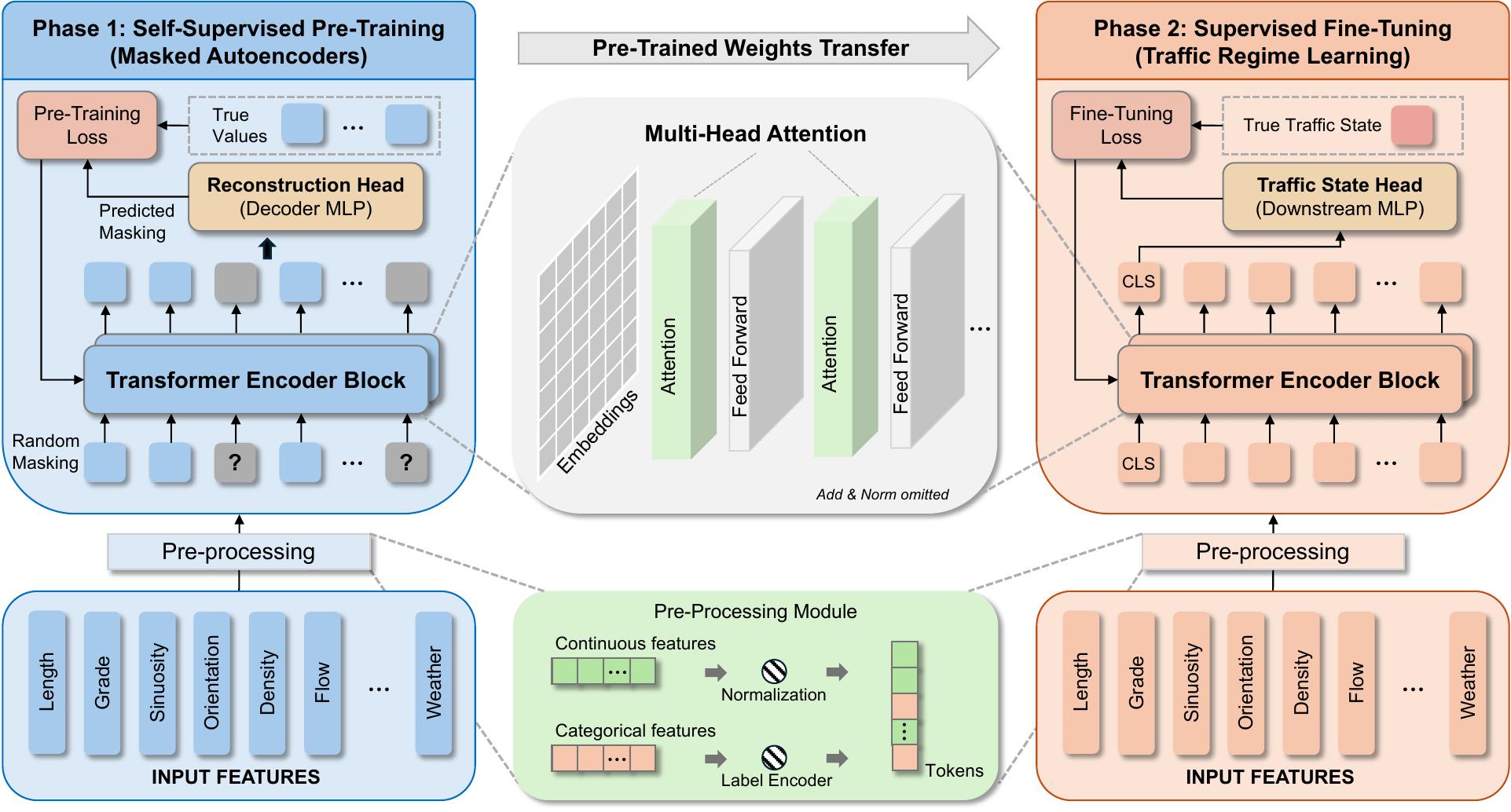}
\caption{\textbf{Foundation model learning architecture}. MLP: Multilayer perceptron}\label{fig: framework}
\end{figure}

Fig.~\ref{fig: framework} illustrates this two-phase learning framework. More technical details are provided in Section ~\ref{sec:methods}. In the pre-training phase, a Masked Autoencoder (MAE) Transformer architecture \cite{he2022masked, vaswani2017attention} is employed. Specifically, we mask a random subset of the input features for each data sample and train the model to reconstruct the missing information. A deep Transformer encoder takes the data with random masking as input and maps it into a high-dimensional latent representation. A lightweight decoder then processes this latent representation, alongside specialized mask tokens, to reconstruct the original input vector.

By applying this framework, the model can pre-train on large archives of historical, unlabeled data. Through the process of masking and reconstructing specific features, the Transformer learns a deep, holistic understanding of the underlying semantics, physical constraints, and network structure of the road links without requiring explicit supervision. Once this foundational representation of the road links, traffic states, and environmental context is learned, the lightweight reconstruction head is discarded. The robust, pre-trained encoder weights are retained and efficiently fine-tuned on task-specific, labeled datasets. This fine-tuning process appends a specialized prediction head for estimating traffic state. Ultimately, this two-phase approach produces a foundation model capable of potentially internalizing transferable traffic-flow physics and the complex realities of the physical world. 

\subsection{Scope, Objectives, and Contributions}

The pre-training and fine-tuning pipeline is a standard practice in foundation model development. Pre-training enables the model to learn generalizable latent representations through self-supervised learning without requiring task-specific labels. These representations can capture underlying relationships among heterogeneous input features, reduce the amount of labeled data and computational effort required for downstream tasks, and facilitate transferability across tasks and geographical regions. 

This study focuses on one downstream application as an illustrative example: characterizing traffic flow regimes through link-level traffic state estimation. Specifically, we first use travel time estimation with focal link inputs as the primary downstream task and extend the analysis in different perspectives, including (1) separate estimation tasks for traffic flow, density, and speed; (2) consideration of the dynamics of neighboring links; (3) traffic forecasts for future horizons, to provide a more comprehensive illustration of the proposed model framework.

The detailed network GIS data rely on transportation network models maintained by state agencies, while probe data are usually obtained from commercial providers. Due to data availability constraints, we were only able to obtain relevant datasets for several counties in Southern California, USA. Therefore, although we propose a learning framework for foundation model development, the present implementation should be interpreted as a proof-of-concept study using the largest integrated multi-source road-link dataset available to us, rather than as a claim of exhaustive global-scale pre-training. We use the term \textit{foundation-model framework} to refer to the proposed reusable pre-training and fine-tuning framework for learning transferable traffic flow representations, and the term \textit{foundation-model prototype} to refer to the specific model trained using the available datasets.

Motivated by recent advances in data availability and AI models, and addressing the aforementioned research gaps, the core objectives and contributions of this research are fivefold: (1) to move beyond conventional traffic flow models that primarily rely on traffic flow characteristics by integrating a rich set of multi-source features associated with traffic and the built and natural environments into a unified fully-differentiable machine learning framework; (2) to establish a scalable, two-phase learning framework (pre-training and fine-tuning) for road links that yields a reusable pre-trained model for various downstream transportation planning and operations applications; (3) to learn a robust, high-dimensional latent representation that captures underlying real-world traffic flow physics at the macroscopic or mesoscopic level and demonstrates transferability across geographical regions, serving as a proof of concept for the proposed foundation-model framework; and (4) to analyze the learned high-dimensional latent representations and uncover data-driven insights that validate, complement, or extend existing traffic flow theories and practices.

\section{Results}\label{sec:results}

The model is trained using data from Los Angeles County (LA) from November 2025 and validated using data from December 2025. The trained model is further evaluated on a testing set from Los Angeles County, as well as on data from two additional counties: Orange County and San Diego County. Evaluation on these two external counties constitutes a zero-shot transfer setting. The temporal resolution of the dataset is 5 minutes. A more detailed description of the datasets is provided in Section \ref{sec:methods}.

\begin{figure}[!htbp]
    \centering
    \begin{subfigure}[t]{\textwidth}
        \caption{}
        \label{fig:convergence}
        \centering
        \includegraphics[width=0.99\linewidth]{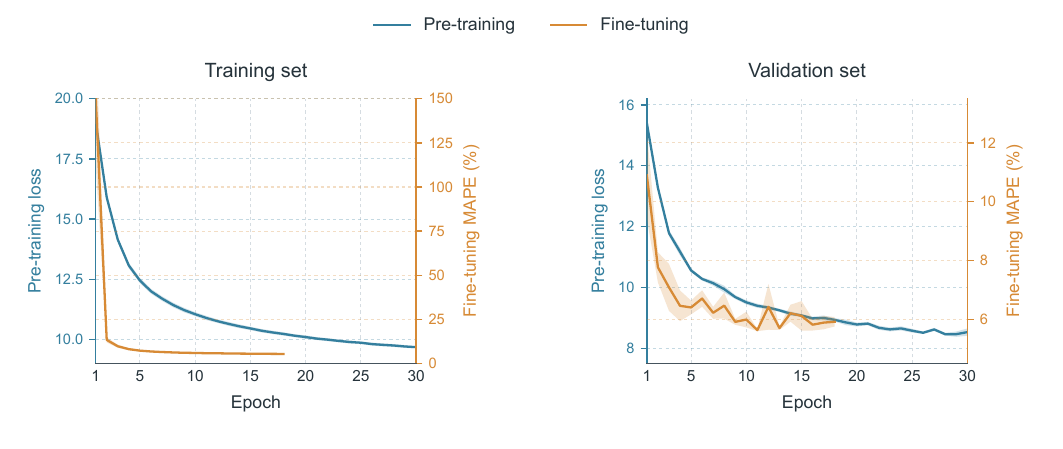}
    \end{subfigure}
    \par
    \begin{subfigure}[t]{0.53\textwidth}
        \caption{}
        \label{fig:temporal_error}

        \centering
        \includegraphics[width=\linewidth]{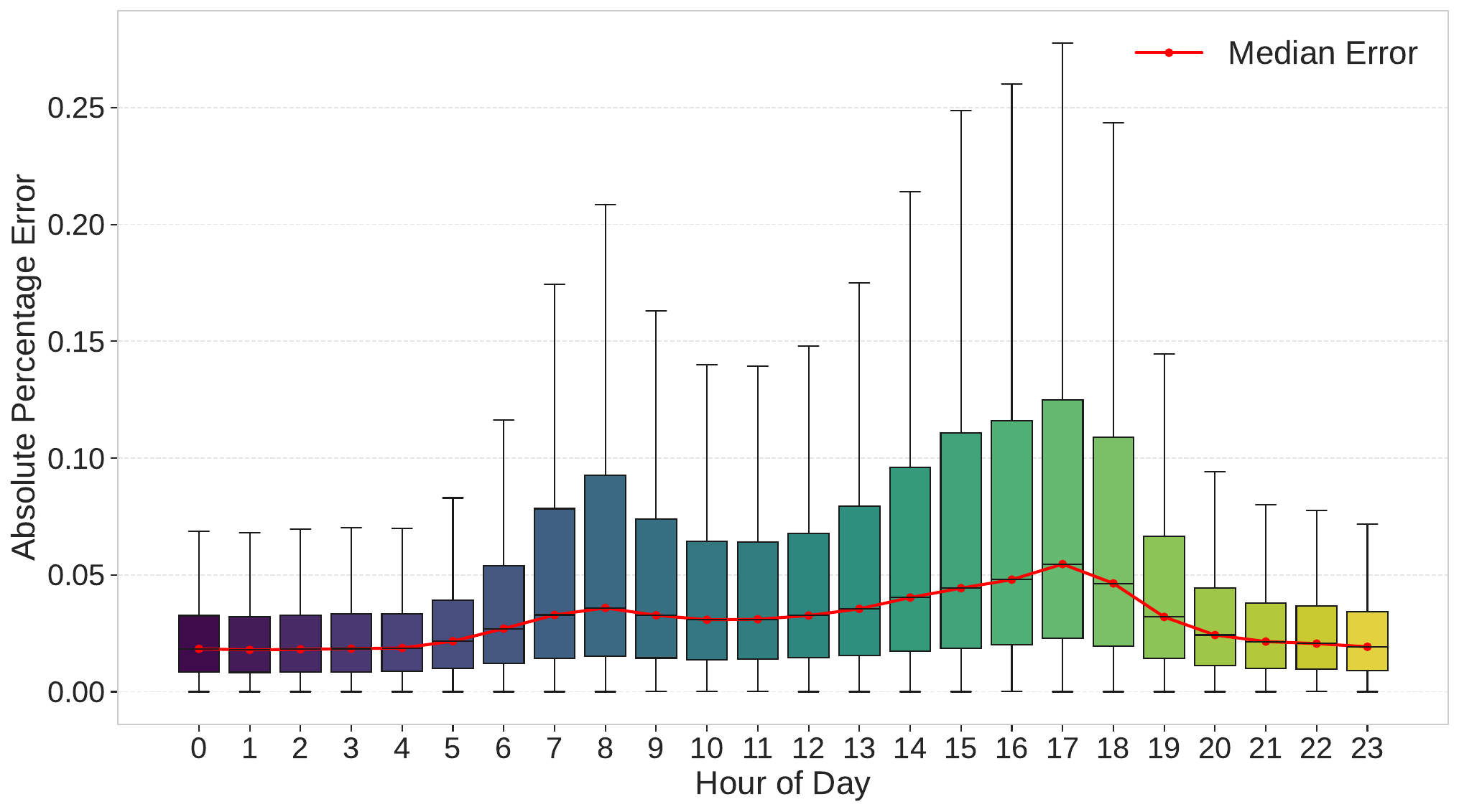}
    \end{subfigure}
    \hfill
    \begin{subfigure}[t]{0.46\textwidth}
        \caption{}
        \label{fig:feature_importance}

        \centering
        \includegraphics[width=\linewidth]{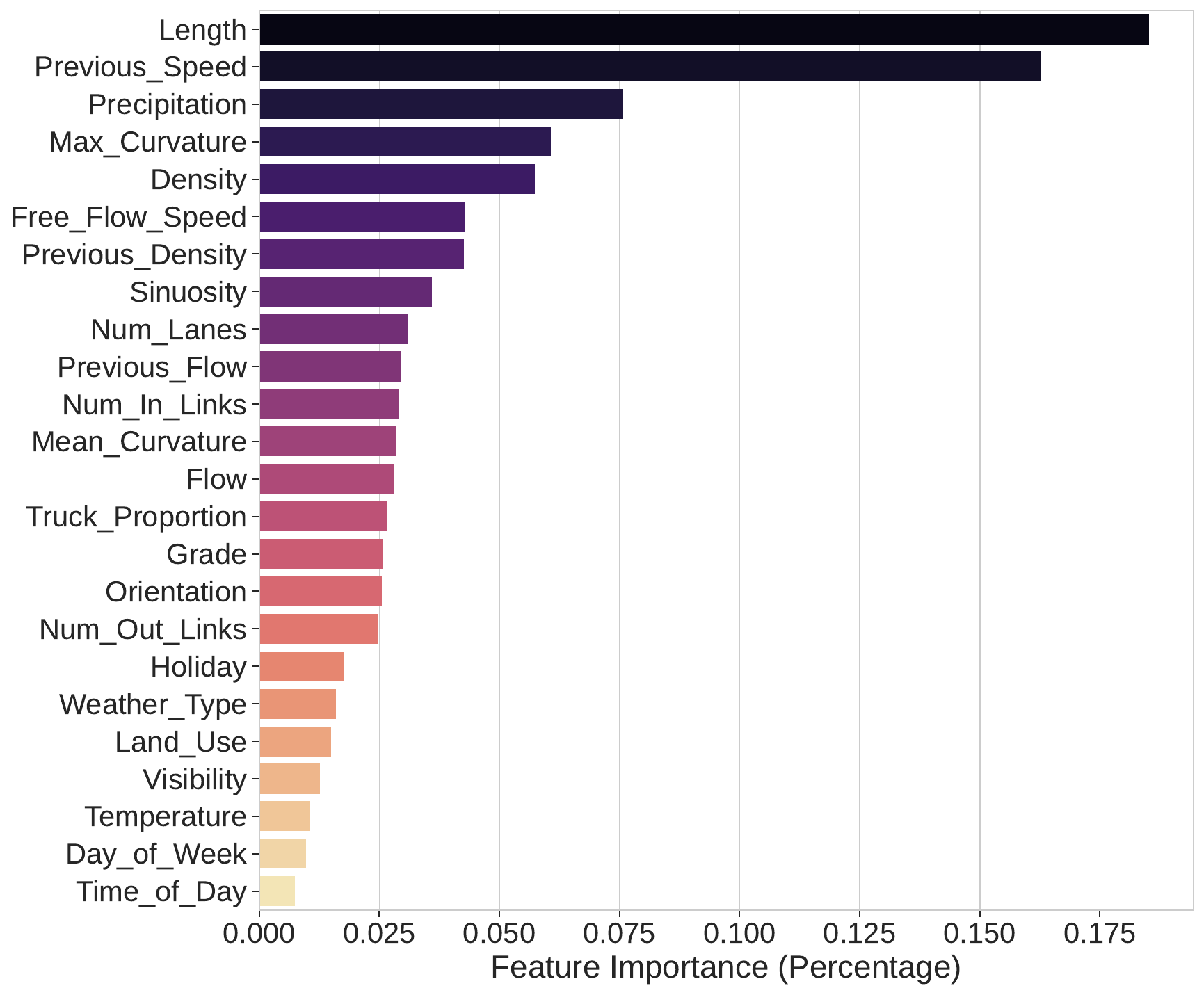}
    \end{subfigure}
    \par
    \begin{subfigure}[t]{\textwidth}
        \caption{}
        \label{fig: w/o pretraining}
        \centering
        \includegraphics[width=0.99\linewidth]{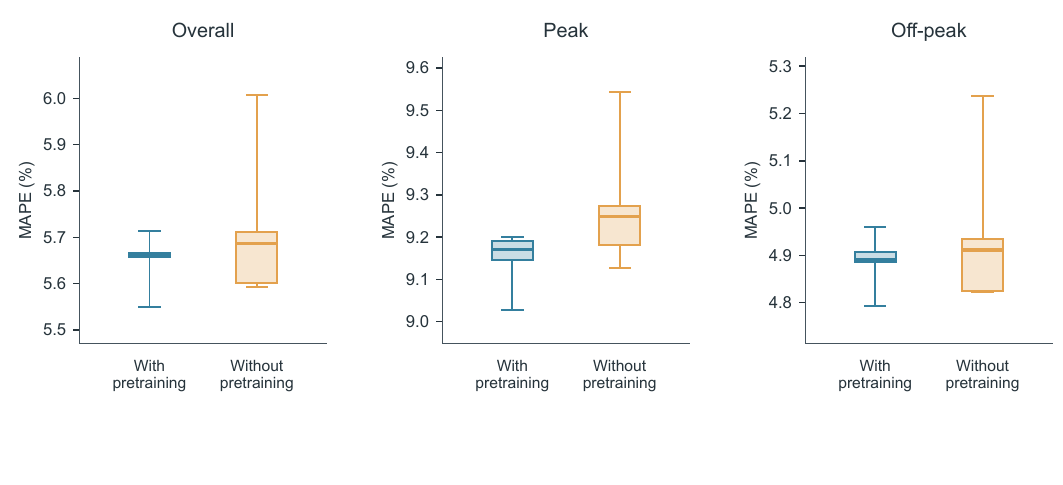}
    \end{subfigure}
    \caption{\textbf{Learning performance and feature evaluations}. \subref{fig:convergence}. Convergence curves for the self-supervised pre-training phase and the supervised fine-tuning phase, demonstrating effective learning. The shaded bands represent variation across different random seeds. \subref{fig:temporal_error}. Temporal distribution of the fine-tuning prediction error across different hours of the day. The model maintains high accuracy, with a slight increase in error variance during daytime peak congestion periods. \subref{fig:feature_importance}. Relative feature importance obtained by gradient-based sensitivity analysis for the downstream travel time learning task. \subref{fig: w/o pretraining}. Effect of pretraining. Panels show travel-time MAPE across all test observations, weekday peak periods and off-peak periods, with different random seeds.}\label{fig: learning results}
\end{figure}

The pre-training and fine-tuning phases are terminated when the loss on validation set fails to improve by at least a predefined convergence threshold ($10^{-3}$) over a consecutive number of epochs (10 in this study). We run the experiments using five different random seeds. Fig.~\ref{fig:convergence} illustrates the convergence of training and validation loss during both the pre-training and fine-tuning phases.  The fine-tuned model achieves an average Mean Absolute Percentage Error (MAPE) of 5.6\% of travel time estimation on the validation set. Fig.~\ref{fig:temporal_error} presents the variation of absolute percentage error by hour of the day, indicating higher variability during the morning and afternoon peak periods. Overall, these results demonstrate the effectiveness of the model.

Although pretraining is widely recognized as a valuable component of foundation model development for large-scale downstream tasks, we conducted experiments with and without pretraining to examine its effects on our prototype datasets. In Fig.~\ref{fig: w/o pretraining}, “With pretraining” denotes self-supervised pretraining followed by supervised fine-tuning, as described previously; “Without pretraining” denotes supervised training from scratch with random initialization, using the same architecture, input features, and prediction head. The results indicate that pretraining provides a slight improvement in travel-time estimation, with a larger improvement during peak periods. Pretraining also reduces variability across training runs. These findings provide preliminary evidence of the benefits of pretraining within the limited scope of our prototype datasets. Previous studies \cite{yoon2020vime, erhan2010does} offer potential explanations for these improvements: In the present study, masked-feature reconstruction may encourage the model to learn dependencies among static, dynamic, and contextual features before supervised training. The resulting latent representations may provide a useful initialization for downstream tasks and reduce sensitivity to training randomness.

\subsection{Travel Time Influencing Factors Extend Beyond Flow Dynamics}

Fig.~\ref{fig:feature_importance} details the relative importance of the input features in the fine-tuned model. For detailed description of each data feature, please refer to Appendix. It is not surprising that Link Length emerges as the most critical determinant of travel time, followed closely by the dynamic feature represented by Previous Speed (speed at the previous time interval in this set of experiments). Notably, Precipitation ranks as the third most important feature overall. This underscores the model’s capability to fuse heterogeneous data and correctly identify environmental disruptions as important drivers of travel time variance. Furthermore, geometric features such as Max Curvature and Sinuosity rank highly, validating the premise that physical road design also largely influences vehicle kinematics and travel time. Interestingly, purely temporal categorical variables (Day of Week, Time of Day) rank at the bottom; this suggests that the model effectively deduces the temporal context directly from the continuous dynamic variables (e.g., current density and previous speed) rather than relying on explicit time labels. In other words, traffic flow dynamics vary by time of day and day of week, but the mechanism of how traffic flow interacts with built and natural environment has little to do with the pure labels of time and date.

\subsection{Fine-Tuned Model Learns the Underlying Physics of Traffic Regime}

\begin{figure}[!htbp]
    \centering
    \begin{minipage}[t]{0.66\textwidth}
        \vspace{0pt}
        \centering

        \begin{subfigure}[t]{\linewidth}
            \centering
            \caption{}
            \label{fig:latent_clusters}
            \includegraphics[width=\linewidth]
                {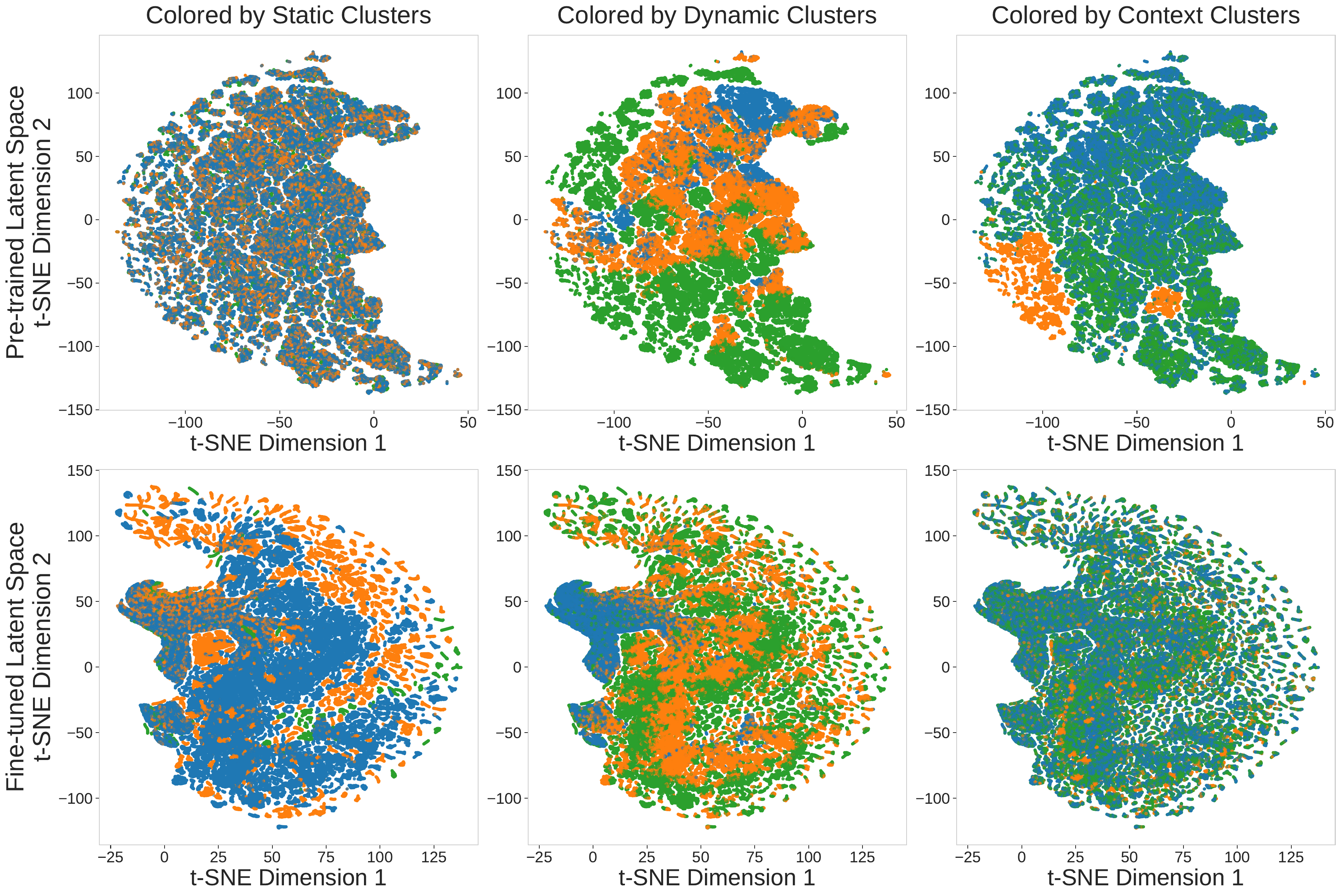}
        \end{subfigure}

        \par\medskip

        \begin{subfigure}[t]{0.65\linewidth}
            \centering
            \caption{}
            \label{fig:latent_congestion}
            \includegraphics[width=\linewidth]
                {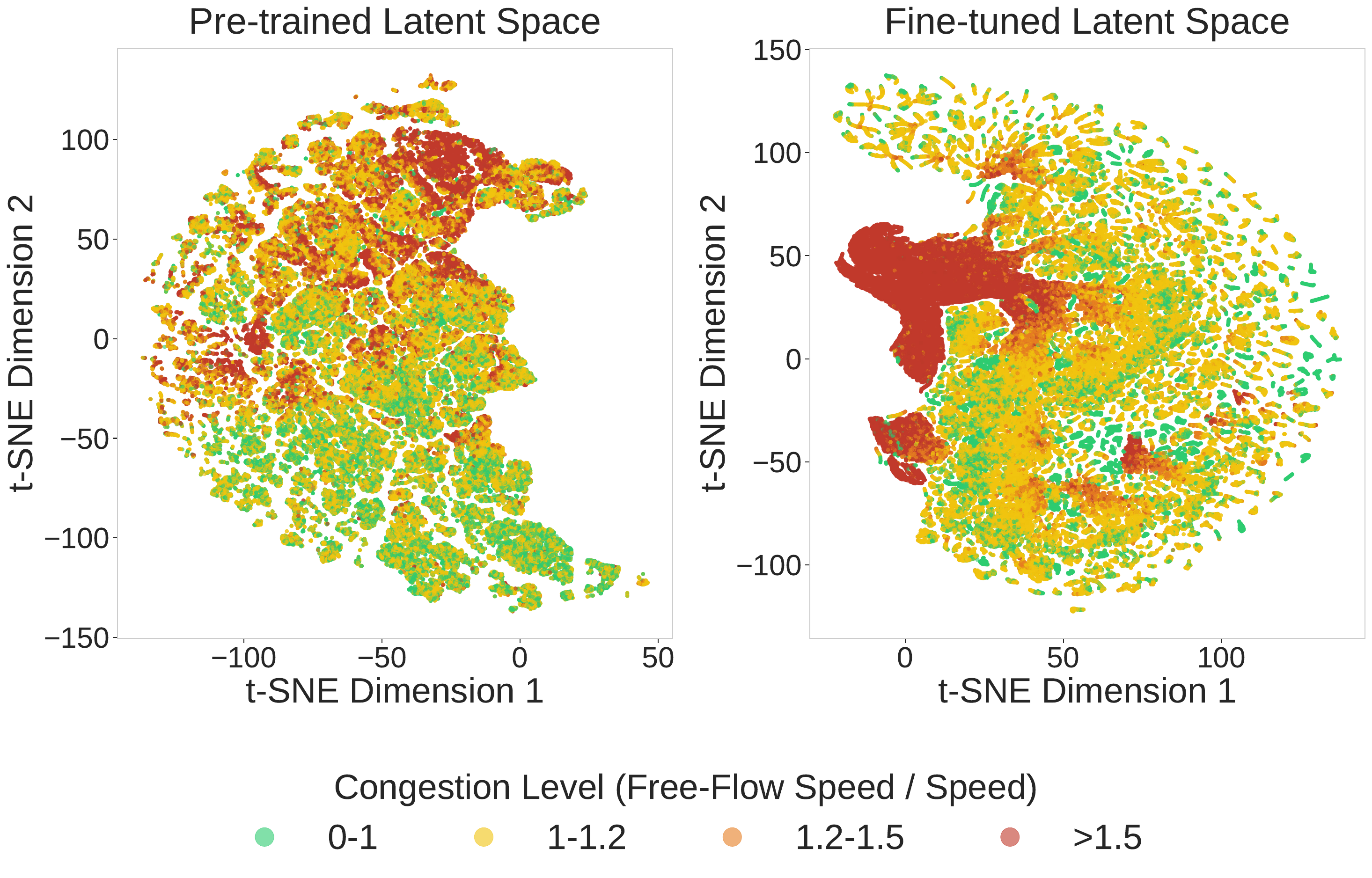}
        \end{subfigure}
        \hfill
        \begin{subfigure}[t]{0.34\linewidth}
            \centering
            \caption{}
            \label{fig:embedding_shift}
            \vspace{6pt}
            \includegraphics[width=\linewidth]
                {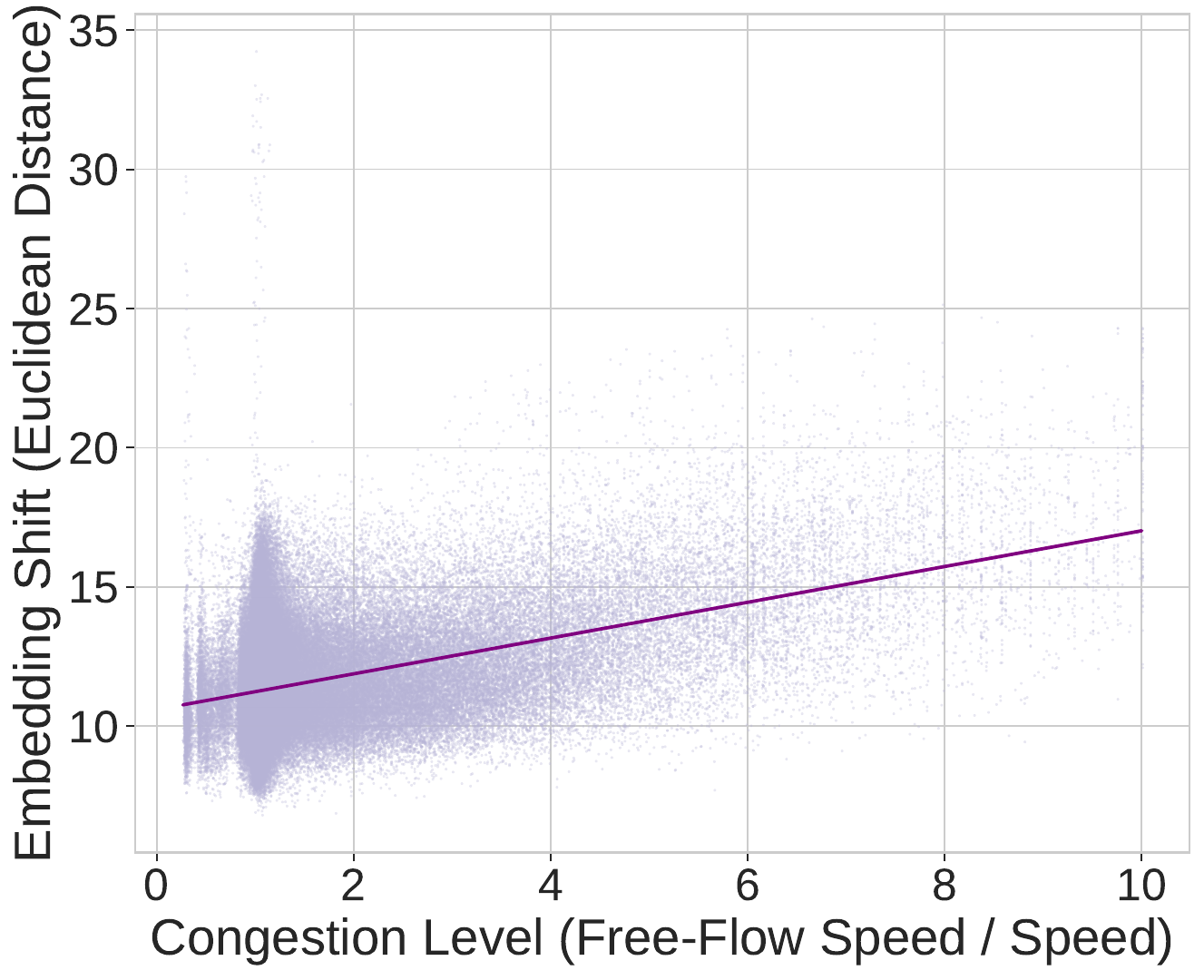}
        \end{subfigure}

    \end{minipage}
    \hfill
    \begin{minipage}[t]{0.33\textwidth}
        \vspace{0pt}
        \centering

        \begin{subfigure}[t]{\linewidth}
            \centering
            \caption{}
            \label{fig:latent_physics}
            \includegraphics[width=\linewidth]
                {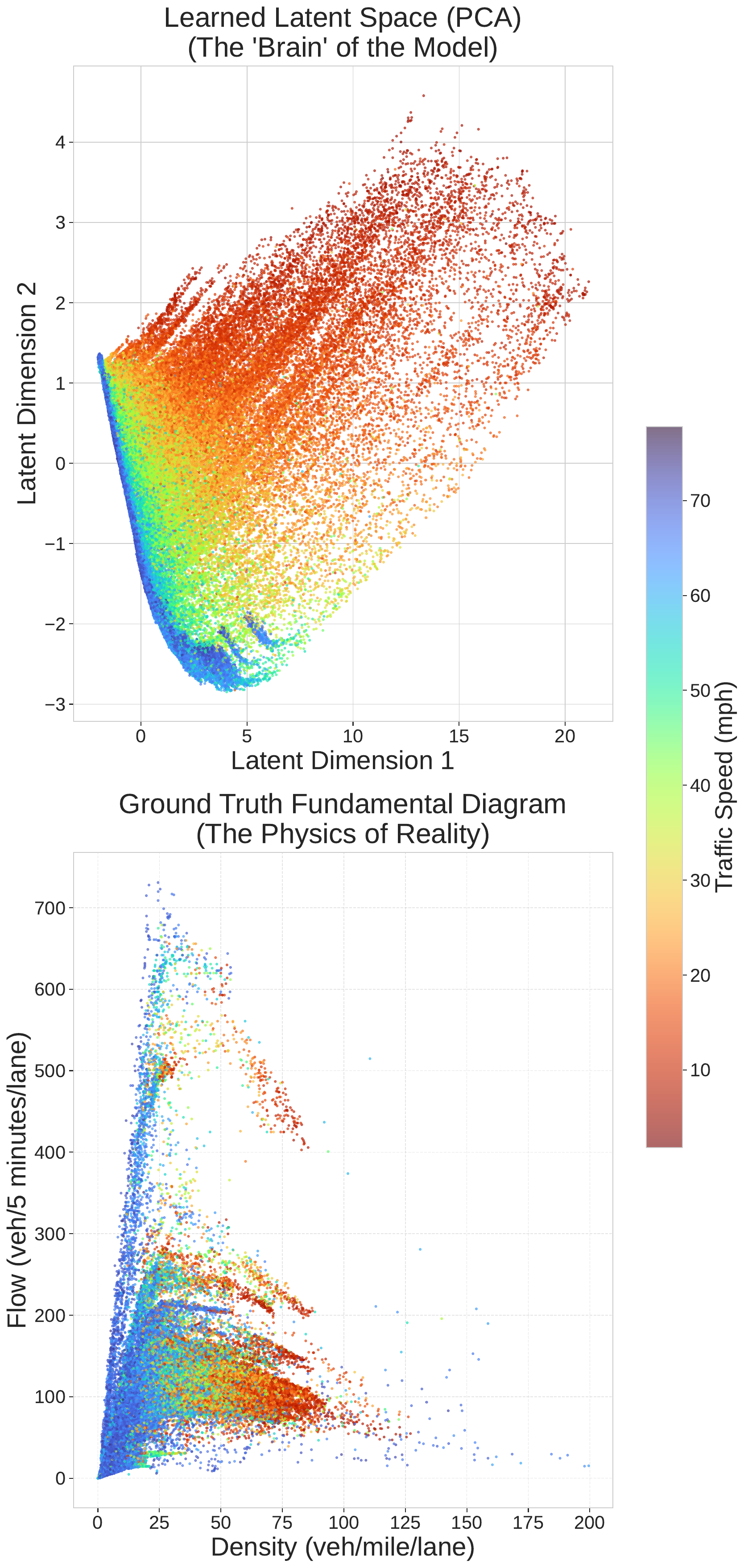}
        \end{subfigure}

    \end{minipage}

    \caption{\textbf{Fine-tuned model learns the underlying physics of traffic regime}. Each dot corresponds to a data sample. 300 K random samples are used for visualization. The dimension of embeddings is 64 in this study. \subref{fig:latent_clusters}. Two-dimensional t-SNE projections of the high-dimensional latent space representations generated by the pre-trained model (top row) and the task-specific fine-tuned model (bottom row). Embeddings are colored according to metadata clusters representing static geometric infrastructure (left), dynamic physical traffic states (center), and environmental/temporal context (right). \subref{fig:latent_congestion}. The t-SNE projections colored by the continuous congestion level (defined as the ratio of free-flow speed to observed speed). While the pre-trained latent space (left) exhibits dispersed congestion characteristics, the supervised fine-tuning objective strictly stratifies the latent space (right), forcing highly congested states (red) into densely concentrated geometric regions completely isolated from others. \subref{fig:embedding_shift}. The plot maps the structural embedding shift (measured as the Euclidean distance between a link state's pre-trained embedding and its corresponding fine-tuned embedding) against its observed congestion level. It shows positive correlation. \subref{fig:latent_physics}. Organic discovery of the fundamental diagram. The top panel displays a PCA projection of the fine-tuned latent space, representing the internal ``brain" of the model. The bottom panel illustrates the empirical ground truth fundamental diagram mapping traffic flow against density. Both manifolds are colored by observed traffic speed. The foundation-model prototype discovers and learns the fundamental physical laws governing vehicular traffic in a high-dimensional latent space.}\label{fig: physics}
\end{figure}

To understand how the model's internal representations (learned embeddings for each data sample) adapt to the downstream prediction task, we visualize the high-dimensional latent spaces (64 dimensions in this study) using t-SNE (t-distributed Stochastic Neighbor Embedding) projections (Fig.~\ref{fig:latent_clusters} and \ref{fig:latent_congestion}. t-SNE is a non-linear dimensionality reduction technique used to visualize complex, high-dimensional data by projecting it into low-dimensional (e.g., 2D) space \cite{van2008visualizing, hinton2002stochastic}. It ensures that nearby points in high-dimensional space remain close together. 

In Fig.~\ref{fig:latent_clusters}, different colors represent different groups  categorized by t-SNE, and the legends are omitted for conciseness. The top row of Fig.~\ref{fig:latent_clusters} illustrates the pre-trained embeddings, which successfully aggregates traffic states into cohesive groups based on general dynamic and contextual similarities. However, static road attributes (e.g., lanes, length, sinuosity, etc.) show weak differentiation. We attribute this to the nature of the pre-training task using masked autoencoders: static features are temporally invariant and exhibit weaker correlations with dynamic features and contextual conditions, making them more difficult to reconstruct when masked and thus leading the model less capable to recognize the distinct representation of static features.

Conversely, based on the second row of Fig.~\ref{fig:latent_clusters}, the fine-tuned model shows improvements in static feature groups. This indicates that accurate travel time learning requires disentangling the physical features of the infrastructure, which makes perfect sense. It also shows that the fine-tuned model exhibits a degradation in contextual features (e.g., time, day). This suggests an information redundancy effect: for instantaneous travel time prediction, explicit static and dynamic features provide stronger signals of traffic conditions, making certain contextual features with overlapping information less dominant and less distinguishable under the single objective of travel time learning. In addition, contextual features alone are insufficient to differentiate traffic regimes; rather, they need to be jointly considered with static and dynamic features to effectively characterize traffic conditions. Note that projecting the high-dimensional latent space into two dimensions for visualization may underrepresent the true extent of clustering. Therefore, our focus here is on the change from the pre-trained model to the fine-tuned model, rather than the specific grouping patterns.

Different downstream tasks may drive the fine-tuned embeddings towards different level of priority for different features. Fig.~\ref{fig:latent_congestion} illustrates the projected latent space, where the embeddings are colored by congestion level. The results clearly reveal a functional shift induced by the fine-tuning task for traffic regime learning. While the pre-trained latent space exhibits some dispersion, the fine-tuned space becomes highly stratified. The supervised travel time objective forces the network to explicitly disentangle uncongested states from congested states, grouping them into distinct manifolds. 

Fig.~\ref{fig:embedding_shift} maps the magnitude of the embedding shift, measured as the Euclidean distance between a state's pre-trained and fine-tuned embedding vectors against its corresponding congestion level. The results demonstrate a clear positive correlation. During fine-tuning, highly congested states experience substantial shifts in the latent space. This confirms that the model selectively allocates its representational updates to resolve the complex, non-linear traffic dynamics.

Fig.~\ref{fig:latent_physics} shows a validation of the model's capability in learning the traffic physics. The top panel visualizes a Principal Component Analysis (PCA) of the learned latent space, referred to as the ``Brain" of the model. The bottom panel displays the empirical ground-truth Fundamental Diagram \cite{greenshields1935study, lighthill1955kinematic1} mapping flow against density, colored by traffic speed. Strikingly, without any explicit programmatic constraints or hard-coded traffic flow equations, the neural network's learned latent space organically forms a geometric manifold that not only mirrors the traditional fundamental diagram but also appears smoother and more structured. The latent dimensions naturally discover and replicate the physical reality of the network: the free-flow states gradually transition into the congested regime. It is coherent with the fact in producing traditional fundamental diagrams that there is a wide spectrum of flow, density and speed values at any point of a fundamental diagram due to complex relations between traffic and environment. However, in the latent space, the pattern appears more clear and well defined even with merely the first two dimensions, particularly when the traffic speed is greater than 30 mph. In the cases where traffic speed is low, namely congestion is heavy, the clustering is less concentrated, indicating many other factors are likely in play to determine flow dynamics. This structural alignment provides clear evidence that the model internalizes the fundamental physical laws governing traffic regime, by considering a comprehensive set of static, dynamic, and contextual features and leveraging advanced learning technologies.

\subsection{Foundation-Model Prototype Exhibits Zero-Shot Transferability}

To rigorously evaluate the performance and robustness of the fine-tuned model, we benchmark it against three representative classes of baseline models: the Greenshields model \cite{greenshields1935study}, a classical traffic flow model, i.e., assuming a linear relationship between speed and density and that ``flow = speed \(\times\) density" holds; a highly optimized tree-based ensemble (LightGBM); and a standard deep Multi-Layer Perceptron (Deep MLP). For the Greenshields model, parameters are first calibrated individually for each road link in the training set to capture link-specific characteristics, after which travel time estimation is performed. For both LightGBM and Deep MLP, the same input feature set used to train the foundation model is adopted to ensure a fair comparison. All machine learning models are tuned using the Optuna framework \cite{akiba2019optuna}, an automated hyperparameter optimization tool for machine learning.

\begin{figure}[!htbp]
    \centering

    \begin{subfigure}[t]{0.61\textwidth}
        \centering
        \caption{}
        \label{fig:model_comparison}
        \includegraphics[width=\linewidth]
            {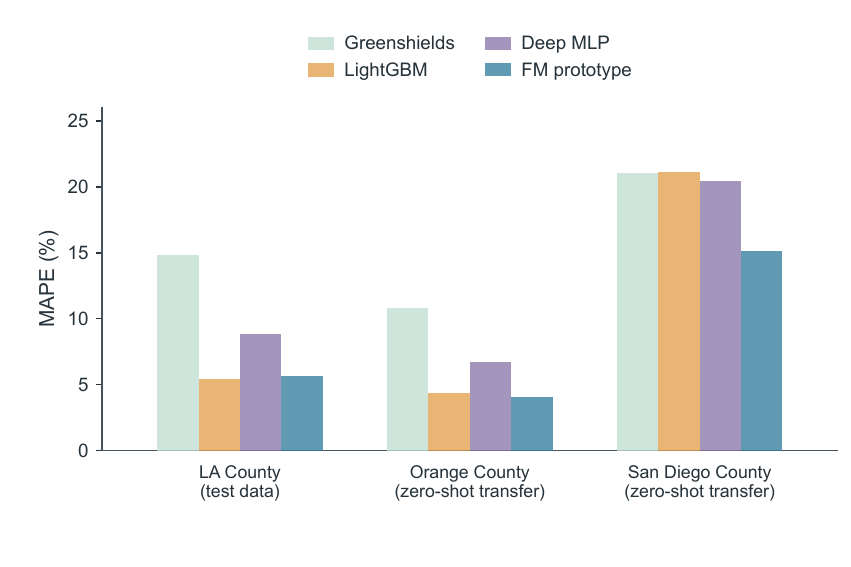}
    \end{subfigure}
    \hfill
    \begin{subfigure}[t]{0.38\textwidth}
        \centering
        \caption{}
        \label{fig:spatial_transfer}
        \includegraphics[width=0.9\linewidth]
            {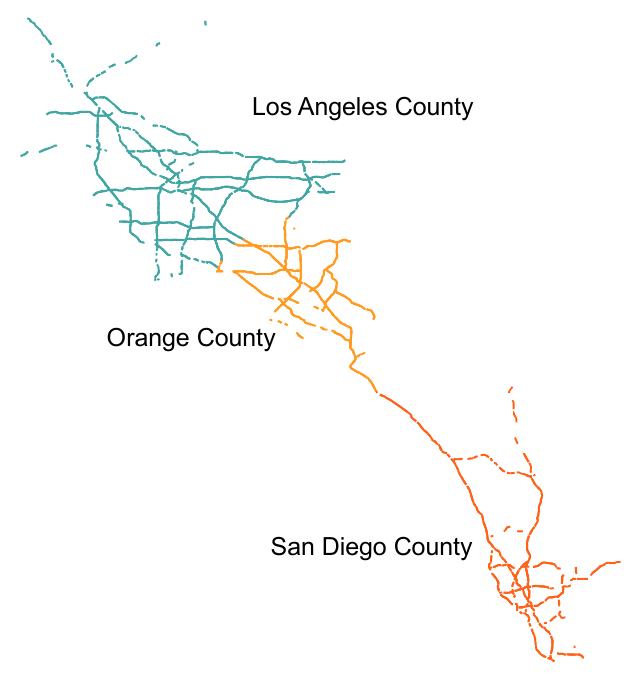}
    \end{subfigure}

    \caption{\textbf{Model comparisons and zero-shot spatial transferability of the Foundation-Model prototype (FM prototype)}. \subref{fig:model_comparison}. The bar chart evaluates the MAPE of the proposed model against a traditional traffic flow model (Greenshields), a tree-based machine learning model (LightGBM), and a standard neural network (Deep MLP). Performance is reported across three geographic domains: the test set from the training region (LA County), and two unseen zero-shot transfer regions (Orange County and San Diego County). \subref{fig:spatial_transfer}. The geometry of road links of LA County, Orange County, and San Diego County.}\label{fig: comparison}
\end{figure}

Fig. ~\ref{fig:model_comparison} shows the mean MAPE values across training runs with different random seeds. When evaluated on the test dataset from the training region (LA County), the purely data-driven machine learning models outperform the traditional Greenshields model. Both the LightGBM and the Foundation-Model prototype achieve high accuracy, effectively capturing the complex non-linearities of the traffic. The Deep MLP lags behind, indicating that standard feed-forward networks struggle to process heterogeneous (continuous and categorical) inputs and learn the underlying physics as effectively as tree-based or attention-based architectures.

The true test of a foundation model is its ability to generalize beyond its training distribution. We evaluate the model, trained exclusively on LA County data, on two completely unseen regions: Orange County and San Diego County, without any fine-tuning or parameter updates (zero-shot transfer). It is worth noting that LA and Orange Counties are geographically adjacent and fall within the jurisdiction of the same metropolitan planning organization (MPO), the Southern California Association of Governments (SCAG). Consequently, they are more likely to share similar standards and practices for transportation network modeling, GIS data representation, data collection, and preprocessing. In contrast, San Diego County is geographically farther from the training region (LA County) and falls within the jurisdiction of a different metropolitan planning organization, the San Diego Association of Governments (SANDAG).

This introduces potential differences in network modeling conventions, GIS data structures, data processing pipelines, regional planning practices, and underlying traffic characteristics. Therefore, while both Orange and San Diego Counties represent zero-shot spatial transfer scenarios, the latter provides a more stringent test of the model's generalizability across geographical and institutional boundaries. 

The results of Orange County on different models show a similar pattern to those of the LA County test set. However, the San Diego County evaluation reveals a clear shift. In this case, the limitations of traditional machine learning become more apparent. The LightGBM model suffers performance degradation, which indicates that the tree-based ensemble may overfit to the specific, regional features of LA County and failed to learn generalizable rules. Conversely, the Foundation-Model prototype exhibits robustness, maintaining a much lower error. This indicates that it internalizes transferable traffic-flow structure rather than merely memorizing local geographic or dynamic patterns, allowing it to adapt effectively to entirely unseen network topologies and traffic circumstances.

Although the Foundation-Model prototype achieves predictive accuracy comparable to that of LightGBM on the LA County test dataset in this experiment (we have other experiments presented later), its primary advantage lies in its spatial transferability. The objective of foundation model development is not necessarily to maximize predictive accuracy within a specific region, but to learn generalizable representations that remain effective across diverse geographical regions and traffic conditions. Sophisticated machine learning models may achieve higher accuracy within the training region by capturing region-specific patterns, yet struggle to generalize to unseen regions. More advanced architectures, such as graph neural networks (GNNs), often explicitly encode region-specific network topology and spatial dependencies, limiting their direct applicability to unseen transportation networks without additional adaptation or retraining. From this perspective, a model with comparable or even marginally lower in-region accuracy but stronger transferability is more desirable for large-scale transportation applications. Such transferability reduces reliance on region-specific training data and model calibration, facilitating broader deployment across heterogeneous transportation networks.

\subsection{Learn Structured Representations of Traffic States Without Past Dynamics Inputs}

We further adapt the Foundation-Model prototype to three additional tasks beyond travel time estimation: estimating the contemporaneous speed, flow, and density of road links. The objective is to evaluate traffic state estimation when past observations of the target variable are unavailable as inputs. Unlike the original setting, where Previous Speed is included as an input feature, the experiments in this section exclude the historical information (i.e., the value in any previous time intervals) of the traffic variable being estimated. Specifically, we train a separate model for each target (i.e., speed, flow, and density). In each task, the value of the target variable at the preceding time intervals is excluded from the inputs, while the other two traffic variables remain available, together with the original static road attributes and contextual features. That is, speed estimation uses flow and density (without Previous Speed); flow estimation uses speed and density (without Previous Flow); and density estimation uses speed and flow (without Previous Density).

We reuse the same pre-trained encoder for these new experiments and fine-tune it separately for each downstream task. We also retrain LightGBM and Deep MLP for comparison.

\begin{figure}[!htbp]
\centering
\includegraphics[width=0.85\textwidth]{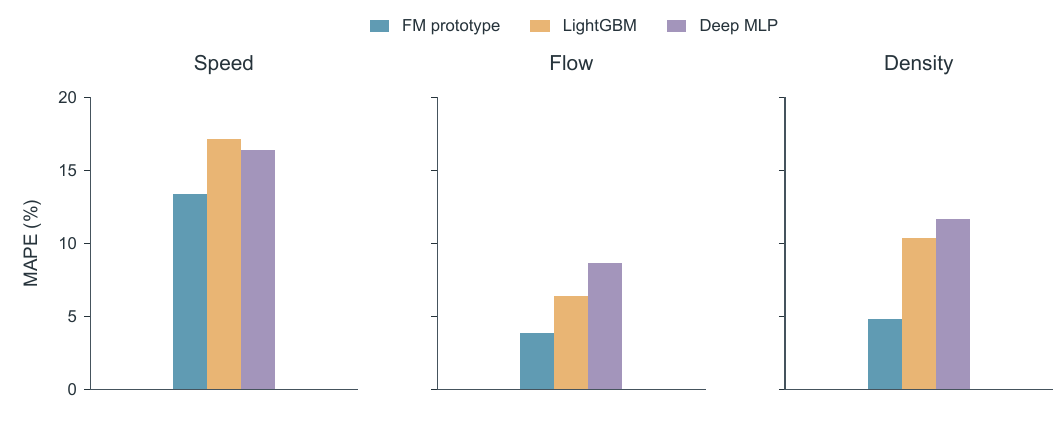}
\caption{
\textbf{Traffic state estimation without input of past dynamics.}
The Foundation-Model prototype, LightGBM, and Deep MLP are compared across speed, flow, and density estimation tasks. Bars show the arithmetic mean of MAPE across training runs with different random seeds. All experiments use the same training, validation, and test datasets.
}
\label{fig:n8-state-estimation}
\end{figure}

\begin{figure}[!htbp]
\centering
\includegraphics[width=\textwidth]{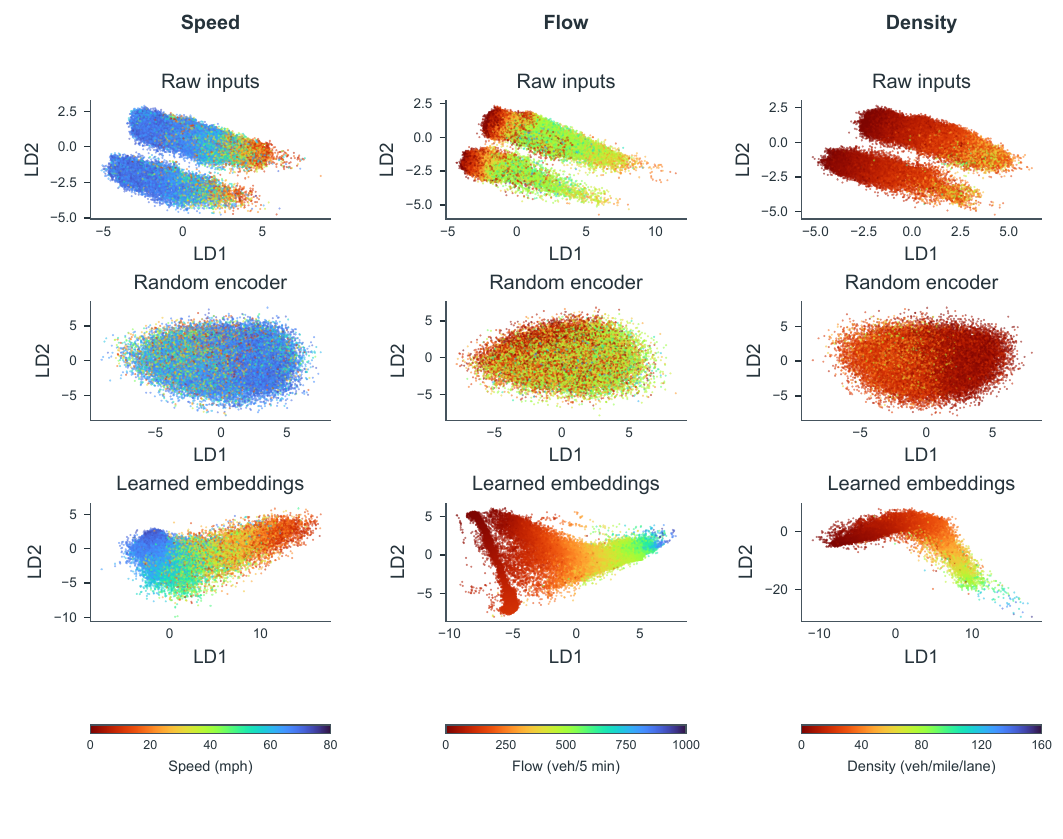}
\caption{
\textbf{Latent representations for separate traffic state estimation tasks.}
Columns correspond to speed, flow, and density estimation targets. Rows show PCA projections of the standardized raw inputs, representations generated by a randomly initialized encoder, and learned embeddings. LD1 and LD2 denote the first two projected dimensions; colors indicate observed target values.
}
\label{fig:n8-state-embeddings}
\end{figure}

Fig.~\ref{fig:n8-state-estimation} shows the mean MAPE across training runs with different random seeds. The Foundation-Model prototype achieves the lowest MAPE for all three targets (speed, flow, and density). These results support the applicability of the proposed framework to multiple traffic state estimation tasks using the same pre-trained model. Comparing with the results in Fig.~\ref{fig:model_comparison}, it suggests that LightGBM may rely heavily on the given information in the previous time interval, potentially limiting its generalizability and predictive performance when such information is unavailable.

To examine the structure of the learned representations, we compare the raw inputs, the representations generated by a randomly initialized encoder, and the learned embeddings, as shown in Fig.~\ref{fig:n8-state-embeddings}. The raw-input projections exhibit coarse color gradients with overlaps among different levels of traffic states, while the randomly initialized encoder produces largely intermixed projections. In contrast, the Foundation-Model prototype reorganizes observations into target-dependent structures characterized by continuous color gradients. For example, the density embeddings form a curved distribution extending from lower- to higher-density observations, while the flow embeddings exhibit a clearly graded arrangement. Such structures are much less evident in the raw-input space, where high-flow and high-density observations are not readily distinguishable from other traffic states. These results provide further evidence that the fine-tuned model captures the underlying physical relationships governing traffic dynamics, forming structured latent representations characterizing a high-dimensional fundamental diagram, as previously discussed in Fig.~\ref{fig:latent_physics}.

\subsection{Foundation-Model Prototype Captures Spatiotemporal Traffic Dynamics}

Traffic dynamics involve spatial dependencies across connected road links and temporal variations in traffic states. To examine whether the proposed model can leverage these spatiotemporal dependencies for short-term prediction, we forecast future speed, flow, and density using current traffic observations from neighboring links. The objective is to assess predictive performance when direct measurements of the focal link's traffic states are unavailable.

In each experiment, the forecast horizons are 5, 10, and 15 minutes, i.e., predicting speed, flow, and density at times $t+5$, $t+10$, and $t+15$ using the information available at time $t$. The inputs at time $t$ comprise the focal link's original static and contextual features, together with upstream and downstream speed, flow, and density measured at time $t$. Both current and historical observations of speed, flow, and density on the focal link are excluded from the inputs for all three prediction targets (i.e., speed, flow, and density). Traffic measurements from neighboring links are averaged separately for the upstream and downstream directions. We also retrain LightGBM and Deep MLP for comparison.

\begin{figure}[!htbp]
    \centering
    \includegraphics[width=\linewidth]{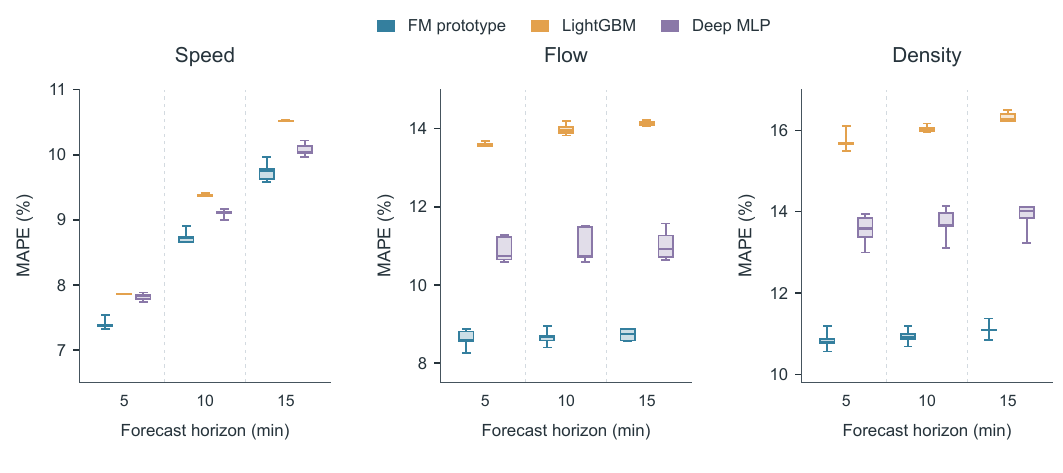}
    \caption{
    \textbf{Traffic state forecasting using information from neighboring links.}
    Box plots show the MAPE distributions for link speed, flow, and density forecasts at horizons of 5, 10, and 15 minutes.
    Both current and historical traffic-state observations of the focal link are excluded from the inputs.
    }
    \label{fig:neighbor-snapshot-multihorizon}
\end{figure}

Fig.~\ref{fig:neighbor-snapshot-multihorizon} shows the prediction performance at the 5-, 10-, and 15-minute forecast horizons. The Foundation-Model prototype achieves the lowest MAPE for all three prediction targets across all forecast horizons. LightGBM performs even worse than DeepMLP. Additionally, prediction errors generally increase with the forecast horizon, reflecting the increasing uncertainty associated with predicting traffic states further into the future.

The results suggest that the learned representations effectively integrate neighboring traffic observations with focal-link attributes to predict traffic states. They demonstrate the prototype's capability to forecast short-term traffic dynamics using spatially connected observations, even in the absence of direct traffic-state measurements on the focal link. More broadly, the findings support the prototype's ability to capture predictive spatiotemporal relationships among connected road links.

\subsection{The Foundation Lies in the Power of Dynamic Attention}

\begin{figure}[!htbp]
    \centering
    \begingroup
        \phantomsubcaption\label{fig:attention_base}
        \phantomsubcaption\label{fig:attention_school}
        \phantomsubcaption\label{fig:attention_truck}
        \phantomsubcaption\label{fig:attention_sinuosity}
    \endgroup
    
    \begin{overpic}[width=0.99\linewidth]{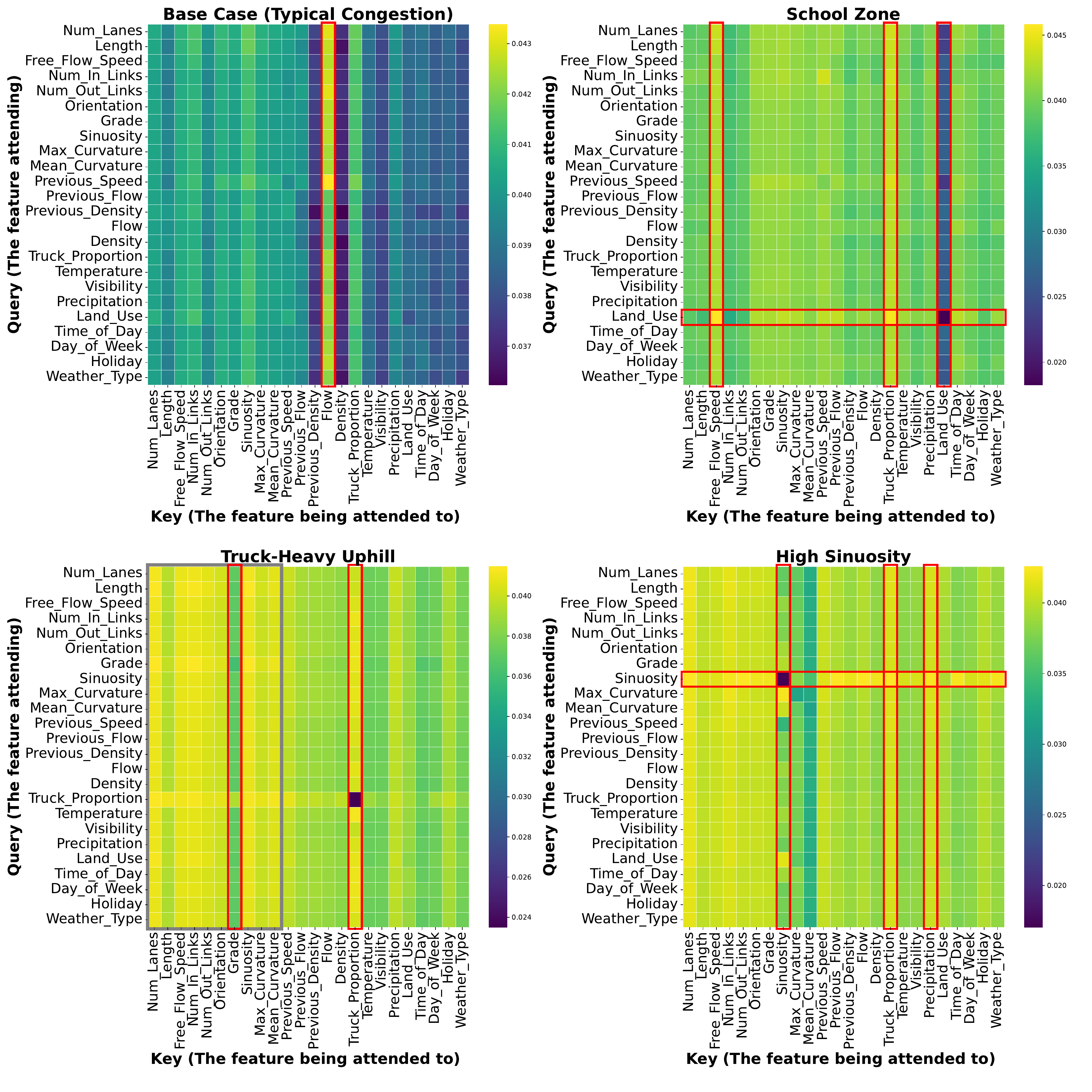}
        \put(1,99){\large\textbf{\subref*{fig:attention_base}.}}
        \put(50,99){\large\textbf{\subref*{fig:attention_school}.}}
        \put(1,48){\large\textbf{\subref*{fig:attention_truck}.}}
        \put(50,48){\large\textbf{\subref*{fig:attention_sinuosity}.}}
    \end{overpic}

    \caption{\textbf{Dynamic attention under varying traffic scenarios}. The heatmaps visualize the average self-attention matrices extracted from the Transformer’s final encoder layer. \subref{fig:attention_base}. A nominal base context of a freeway with typical traffic congestion due to high flow. \subref{fig:attention_school}. An uncongested local road in a urban school zone. \subref{fig:attention_truck}. An uncongested freeway with a 7\% uphill grade and a 70\% truck proportion. \subref{fig:attention_sinuosity}. An uncongested freeway with higher horizontal sinuosity.}
    \label{fig: attention}
\end{figure}

One of the core advantages of the foundation model lies in its multi-head self-attention mechanism, which eschews static parameter weights in favor of dynamic, data-dependent attention. To illustrate this, we extract and visualize the Query-Key attention matrices (average across all heads) produced by the fine-tuned model of travel-time estimation task using different input data samples that represent a baseline condition and three distinct scenarios (Fig.~\ref{fig: attention}). 

Note that the interpretability of attention has been widely debated in the literature \cite{jain2019attention, wiegreffe2019attention, bibal2022attention, clark2019does, brunner2019identifiability, li2023transformers}. The intent of the following analysis is not to interpret specific attention values or draw definitive conclusions about the model’s internal decision-making and the general interpretability of attention. Rather, it aims to illustrate, in a high-level manner, how attention patterns dynamically shift as the primary factors influencing travel time vary across different scenarios.

One of the key strengths of the attention mechanism, which is also a fundamental feature underlying the success of language models, is its context-dependent representation capability. Under different “contexts”, the resulting embedding vector of the same feature value can vary, as the model attends to different features to better encode context-specific meanings. In other words, the formation of embedding vectors dynamically changes across contexts through the routing of attention.

The attention value received by a specific feature reflects the extent to which that feature is attended to when other features form their embeddings to represent certain meanings in the high-dimensional latent space under a given context. Moreover, the learned travel time is parameterized by many other weights and parameters in the neural network. Therefore, attention values should not be directly interpreted as indicators of feature importance with respect to the learning objective. Our intention is not to fully interpret the internal mechanism of this complex black-box model, but rather to illustrate the dynamic shift of attention across different scenarios, which is a foundational strength of attention-based language models and distinguishes them from traditional simple neural network architectures.

The base scenario represents a typical congested condition resulting from high traffic flow on a freeway with a free-flow speed of 70 mph. The roadway is straight and flat, with no truck traffic present. Despite these favorable conditions, congestion reduces the actual travel speed to 40 mph. As shown in Fig.~\ref{fig:attention_base}, the feature \textit{Flow}, indicating heavy traffic, receives the most attention from other features.

The following presents three uncongested scenarios in which travel time is primarily influenced by factors other than the amount of traffic flow.

\subsubsection*{Scenario 1: School Zone (Fig.~\ref{fig:attention_school})}

In this scenario, the geometry is modified to represent a local road within a school zone, characterized by a reduced free-flow speed (25 mph) and a specific land-use designation (Education). Meanwhile, traffic flow and density are reduced to represent uncongested conditions. The resulting attention matrix exhibits a substantial topological shift. The feature \textit{Free-Flow Speed} receives greater attention from other features. Although \textit{Land Use} receives the least attention from other features, it assigns significant attention to \textit{Free-Flow Speed} during the formation of its embedding. This suggests that the learned embeddings encode a traffic state consistent with an uncongested local road with a low free-flow speed. Furthermore, \textit{Truck Proportion} also receives relatively high attention from other features, especially from \textit{Land Use}, indicating that truck presence may also deserve attention in a school zone.

\subsubsection*{Scenario 2: Truck-Heavy Uphill (Fig.~\ref{fig:attention_truck})}

This scenario combines a 7\% uphill grade with a high truck proportion (70\%). Meanwhile, traffic flow and density are reduced to represent uncongested conditions. The results show that \textit{Truck Proportion} receives relatively high attention from other features. \textit{Grade} generally receives less attention from other features, but is exceptionally attended to by \textit{Truck Proportion}. Overall, static features of the built environment receive comparatively higher attention than other feature types, as indicated by the yellow areas on the left. These findings suggest that the model has internalized the underlying physics of truck-heavy uphill conditions, placing greater attention on physical infrastructure. In particular, it identifies that in this scenario a high proportion of heavy vehicles on a steep incline governs travel time, regardless of nominal flow or density.

\subsubsection*{Scenario 3: High Sinuosity (Fig.~\ref{fig:attention_sinuosity})}

In this scenario, the road is modified to exhibit higher \textit{Max (Mean) Curvature} and \textit{Sinuosity}. Meanwhile, traffic flow and density represent uncongested conditions. The results indicate that \textit{Sinuosity} more actively attends to others during embedding formation, trying to refine its high-level meanings in the latent space given such a context. In particular, \textit{Max Curvature} places the strongest attention on \textit{Sinuosity}. Consequently, the learned embeddings encode the interpretation that travel time in this scenario is primarily attributable to greater roadway curvature rather than traffic congestion. Furthermore, in addition to geometric features, \textit{Truck Proportion} and \textit{Precipitation} also receive relatively high attention from other features, indicating that truck presence and precipitation also deserve greater attention under roadway conditions with higher sinuosity. Namely, rainy conditions and truck presence can become critical on roads with high sinuosity. These findings demonstrate that the model captures the impact of geometric constraints, recognizing that even under uncongested conditions with low density, sharper curvature necessitates speed reduction, and that truck presence and weather conditions also require careful attention on highly curved roads.

\vspace{10pt}

The detailed data samples for these scenarios are provided in the Appendix.

\FloatBarrier

\section{Discussions}\label{sec: discussions}

The results presented in this study demonstrate that the proposed foundation-model framework represents a potential shift in the traffic regime learning paradigm by transitioning from rigid, hand-coded physical models to an autonomous, differentiable neural architecture. By leveraging a two-phase learning framework consisting of self-supervised pre-training and supervised fine-tuning, the model captures the fundamental physical laws of traffic kinematics directly from high-dimensional, multi-source data that expand across static, dynamic, and contextual features.

\subsection*{The Multidimensional Nature of Traffic Regime}
One of the key findings of this research is that the factors influencing the traffic state extend beyond those traditionally considered. The model identifies that geometric features, such as grade, sinuosity, and land use, along with other contextual variables such as truck proportion and precipitation, serve as highly influential features. This underscores the foundation model's superior ability to fuse heterogeneous, multi-source data and recognize environmental disruptions and physical road characteristics as potential drivers of traffic regime. 

\subsection*{Discovery of Generalizable Traffic Flow Regime and Zero-Shot Transferability}
The core strength of this approach lies in its ability to discover the true fundamental diagram of traffic flow without explicit programmatic constraints. Unlike traditional machine learning models, which suffered performance degradation when transferred to unseen regions, the foundation-model prototype exhibited higher zero-shot transferability. This supports the interpretation that the attention-based architecture, along with multi-source data, learns transferable traffic-flow structure from features related to traffic and built/natural environment, bridging the long-standing gap between data-driven accuracy and physical generalizability. 

\subsection*{Scalable Differentiability}
Furthermore, the model's end-to-end differentiability (automatically enabled by the pure neural networks) resolves mathematical bottlenecks in large-scale optimization, providing a scalable computational foundation for next-generation intelligent transportation systems (ITS). For example, the model could serve as differentiable state representations or surrogate dynamics within traffic assignment, dynamic routing, and network design.

\subsection*{Pre-trained Link Representations for Diverse Downstream Tasks}
Although traffic state estimation serves as the primary downstream application in this study, the pre-trained dense representations are not inherently restricted to traffic state estimation tasks. In the broader vision of a transportation foundation model trained on large-scale, heterogeneous datasets, the learned embedding summarizes the joint state of a road link, integrating its geometry, traffic conditions, vehicle composition, and environmental context. It can therefore serve as a general-purpose input to various task-specific prediction heads, reducing the need to develop separate models from scratch for individual applications.

Incident detection provides a straightforward example of a classification task. A binary classification head with a sigmoid output could predict the probability of an incident occurring on a road link, using binary cross-entropy as the fine-tuning objective. The learned representations could facilitate the identification of abnormal traffic patterns by accounting for the interactions among traffic dynamics, roadway characteristics, and environmental conditions, rather than relying solely on deviations in individual traffic variables.

Another promising application is to serve as a core computational component of urban digital twins. By continuously integrating real-time observations from heterogeneous data sources, the foundation model could maintain dynamically updated representations of road links across an entire transportation network. These representations could support real-time traffic monitoring, the inference of unobserved traffic states, short-term forecasting, and the identification of emerging disruptions. Furthermore, when coupled with network-level traffic simulation and optimization models, the foundation model could support scenario analysis and the evaluation of alternative operational strategies and policy interventions. Its spatial transferability could also facilitate the development of digital twins across different cities without requiring extensive region-specific model training and calibration.

\subsection*{Future Directions Towards a True Foundation Model}
Expansion to Local Arterials: Currently, the model utilizes PeMS data, which focuses predominantly on major highways and freeways. Future work can include data from local arterials or involve fine-tuning models to capture the unique stop-and-go dynamics and signalized intersections characteristic of urban street networks.

Consideration of Additional Features: With the rapid advancement of ubiquitous infrastructure sensing, connected and automated vehicles, and data-rich mobility technologies from Waymo, Tesla, OEMs, and other providers, future research can further enrich the framework by incorporating additional infrastructure- and vehicle-related features. Examples include 2D and 3D street-view data for roadway geometry, built environment, signage, and lane markings, as well as vehicle composition data capturing electric vehicles, automated vehicles, trucks, and other vehicle classes. These emerging data sources can provide a richer representation of traffic environments and potentially help improve the learning of foundational physics and the model transferability across regions and scenarios.

Large-Scale Global Datasets: This study is intended as proof-of-concept research for a foundation-model framework. The current implementation is constrained by the availability of comprehensive multi-source data, which was limited to Southern California in this study. Therefore, the results should not be interpreted as evidence that the present trained model already exhaustively represents global traffic conditions. Instead, they demonstrate that the proposed pre-training and fine-tuning paradigm can learn reusable representations and transfer across meaningful regional shifts. As comparable data become available for other cities, countries, roadway classes, sensing systems, climates, and driving cultures, a larger foundation model could be trained following the same framework to capture a broader range of global infrastructure designs and mobility behaviors.

\subsection*{Other Future Extensions}

Various Downstream Tasks: As discussed earlier, future research could explore a broader range of ITS applications beyond traffic state estimation, including incident detection, digital twin modeling, and policy evaluation.

Alternative Representation Learning Methods: While this research utilizes a Masked Autoencoder approach for pre-training as an illustrative example, future iterations could explore different representation learning methods, such as contrastive learning or generative adversarial frameworks, to examine which architectures most efficiently encode road segments, dynamic physics, and environmental factors.

Integration with Real-Time Control: Leveraging the fully-differentiable nature, future studies could explore its integration with gradient-based solvers for various tasks such as real-time network routing and dynamic pricing, where simulation and optimization may occur simultaneously.

Generative Models for Downstream Traffic Applications: Another future extension is to move beyond discriminative prediction tasks and train generative models on the learned road-link representations for a broader range of downstream applications. For example, conditional generative models could synthesize plausible future traffic states under alternative weather, demand, incident, control, or infrastructure scenarios, enabling scenario-based planning and stress testing of transportation networks. Diffusion models, variational autoencoders, or autoregressive Transformer decoders could also be built on top of the pre-trained link embeddings to generate temporally coherent trajectories of speed, flow, density, emissions, or queue evolution. Such generative capabilities would potentially make the foundation-model framework useful not only for estimating current traffic conditions, but also for simulating counterfactual network responses, augmenting sparse training data, evaluating rare events, and supporting optimization-based decision making under uncertainty.

\section{Methods}\label{sec:methods}

The methodology framework has been previously shown in Fig.~\ref{fig: framework} and briefly introduced before. Further details are provided below. 

\subsection{Input Representation and Embedding}
The raw data contains both continuous numerical features (e.g., density, flow, free-flow speed) and discrete categorical features (e.g., weather type, land use, time of day). To map these heterogeneous inputs into a unified continuous latent space, we apply feature-specific embedding strategies. We first apply standard normalization to continuous features and numerical label encoding to categorical features, and then map them into the high-dimensional latent space.

Let $\mathbf{x}^{num} \in \mathbb{R}^{F_{num}}$ denote the continuous features and $\mathbf{x}^{cat} \in \mathbb{Z}^{F_{cat}}$ denote the categorical features, where $F_{num}$ and $F_{cat}$ are the total number of numerical and categorical variables, respectively. Continuous features are standard-scaled and projected independently into a $D$-dimensional latent space ($D=64$ in this study) using a single-layer Multi-Layer Perceptron (MLP) with a ReLU activation:
\begin{equation}
    \mathbf{e}_i^{num} = \text{ReLU}(\mathbf{W}_i x_i^{num} + \mathbf{b}_i), \quad \forall i \in \{1, \dots, F_{num}\}
\end{equation}
where $\mathbf{W}_i \in \mathbb{R}^{D \times 1}$ and $\mathbf{b}_i \in \mathbb{R}^D$ are learnable weights. Categorical features are projected using standard lookup-table embeddings:
\begin{equation}
    \mathbf{e}_j^{cat} = \text{Embed}(x_j^{cat}), \quad \forall j \in \{1, \dots, F_{cat}\}
\end{equation}

A special learnable summary token, $\mathbf{e}_{\text{[CLS]}} \in \mathbb{R}^D$, which aggregates the semantics of all tokens, is prepended to the sequence \cite{devlin2019bert}. The final input sequence to the Transformer is constructed by concatenating the [CLS] token and all feature embeddings:
\begin{equation}
    \mathbf{E} = [\mathbf{e}_{\text{[CLS]}}, \mathbf{e}_1^{num}, \dots, \mathbf{e}_{F_{num}}^{num}, \mathbf{e}_1^{cat}, \dots, \mathbf{e}_{F_{cat}}^{cat}] \in \mathbb{R}^{(1 + F_{num} + F_{cat}) \times D}
\end{equation}

\subsection{Self-Supervised Pre-training}
During the pre-training phase, the model learns the underlying correlations among geometric road links, traffic dynamics, and situational context, without requiring labeled travel times. We apply a random masking strategy where a proportion of the input tokens (excluding the [CLS] token) are replaced by a learnable mask token $\mathbf{e}_{\text{[MASK]}}$. The corrupted sequence is processed by a Transformer Encoder. The self-attention mechanism allows the network to dynamically route information and reconstruct the masked features based on the unmasked features. 

Because the inputs are heterogeneous, the reconstruction loss is a composite objective. We apply separate decoding heads to the encoder's output. Given a data sample, let $\mathcal{M}_{num}$ and $\mathcal{M}_{cat}$ denote the indices of the masked numerical and categorical features. The pre-training loss $\mathcal{L}_{pre}$ combines Mean Squared Error (MSE) for continuous variables and Cross-Entropy (CE) for categorical variables:
\begin{equation}
    \mathcal{L}_{pre} = \sum_{i \in \mathcal{M}_{num}} \left\| \hat{x}_i^{num} - x_i^{num} \right\|_2^2 + \sum_{j \in \mathcal{M}_{cat}} \text{CE}(\hat{\mathbf{p}}_j^{cat}, x_j^{cat})
\end{equation}
where $\hat{x}_i^{num}$ is the scalar output of the continuous decoder head, and $\hat{\mathbf{p}}_j^{cat}$ is the logit vector predicted by the categorical decoder head over the unique classes of feature $j$.

\subsection{Supervised Fine-Tuning for Traffic Regime Learning}
Following convergence on the reconstruction task, the decoder heads are discarded, and the pre-trained encoder weights are retained for the downstream task of traffic regime learning. Denote the ground truth traffic state (i.e., travel time, speed, flow, or density, depending on the specific tasks) as $y$. 

During fine-tuning, the full sequence of features and the [CLS] token are processed by the Transformer Encoder. The aggregate  embedding, i.e., the [CLS] token at the final layer, denoted as $\mathbf{z}_{\text{[CLS]}}^{(L)}$, acts as the global, contextualized representation of the road link's state. 

This foundation representation is fed into a downstream MLP prediction head to yield the estimated value $\hat{y}$:
\begin{equation}
    \hat{y} = \mathbf{W}_2(\text{ReLU}(\mathbf{W}_1 \mathbf{z}_{\text{[CLS]}}^{(L)} + \mathbf{b}_1)) + \mathbf{b}_2
\end{equation}

The fine-tuning phase is optimized using the Mean Absolute Percentage Error (MAPE) loss:
\begin{equation}
    \mathcal{L}_{fine} = \frac{1}{N} \sum_{k=1}^N \left| \frac{y_k - \hat{y}_k}{y_k} \right|
\end{equation}
where $N$ is the batch size. 

To prevent overfitting and ensure optimal generalization to unseen data, we employ a patience-based early stopping criterion during both the pre-training and fine-tuning phases. This technique continuously monitors the model's performance on a held-out validation set at the conclusion of each training epoch. If the validation loss fails to improve by at least a predefined convergence threshold for a consecutive number of epochs, the training process is terminated. Then, the model reverts to the checkpoint associated with the lowest validation error, guaranteeing that the final weights represent the model's best predictive capability.

\subsection{Data Sources and Processing}

The model is trained using data from Los Angeles (LA) County. Specifically, the training dataset consists of observations from November 2025, while the validation dataset includes data from the first week of December 2025. The trained foundation model is further evaluated on data from the second week of December 2025 in Los Angeles County, as well as on two additional counties: Orange County and San Diego County. Evaluation on these two external counties constitutes a zero-shot spatial transfer setting, with San Diego County providing a more stringent test of transferability because it falls under the jurisdiction of a different metropolitan planning organization.

The temporal resolution of the dataset is 5 minutes, with a total of 10,042,547 samples in the training set. The data sources are described as follows.

\textit{\textbf{Network GIS}}. The network GIS data provide static attributes of road links, including length, free-flow speed, and number of lanes. Additional features, such as the number of incoming and outgoing links, orientation, sinuosity, and maximum/mean curvature, are derived from the geometric representation of road links. The GIS data for Los Angeles and Orange Counties are obtained from the Southern California Association of Governments, while the data for San Diego County are sourced from the San Diego Regional Data Warehouse.

\textit{\textbf{Land Use Parcel}}. The parcel data include land use classifications (e.g., residential, commercial, industrial, etc.) and parcel geometries. Parcels are spatially matched to road links to determine the dominant land use associated with each link. Data for Los Angeles and Orange Counties are obtained from the Southern California Association of Governments, while data for San Diego County are sourced from the San Diego Regional Data Warehouse.

\textit{\textbf{Digital Elevation Model (DEM)}}. The DEM data provide high-resolution elevation information, which is used to compute the grade of each road link. These data are obtained from the U.S. Geological Survey (USGS).

\textit{\textbf{Traffic Flow and Density}}. Traffic dynamics data, including vehicle counts, occupancy, and truck proportion, are obtained from PeMS (Caltrans). Traffic density is derived from occupancy measurements. PeMS stations are first mapped to the road links defined in the network GIS data. When multiple stations correspond to a single road link, the measurement values are averaged across stations.

\textit{\textbf{Traffic Speed}}. Traffic speed data are obtained from INRIX. INRIX road segments are mapped to the road links defined in the network GIS data. When multiple INRIX segments correspond to a single road link, the harmonic mean speed is computed to represent link-level speed.

\textit{\textbf{Weather Data}}. Weather data, including temperature, weather conditions, precipitation, and visibility, are obtained from the National Oceanic and Atmospheric Administration (NOAA).

Detailed descriptions of all data feature definitions and preprocessing procedures are provided in the Appendix.

\nolinenumbers
\clearpage
\begin{appendices}

\section{Data Feature Definition and Processing}\label{app:data-features}

This section presents the definitions of all data features used in the learning process, along with any preprocessing steps required for their direct use in model training.

\subsection{Data Feature Definition}

The features considered in this study are listed below, together with their definitions and corresponding calculation procedures where applicable.

\subsubsection*{Length}
Length of the road segment. Unit: miles.

\subsubsection*{Number of Lanes}
Number of lanes of the road segment.

\subsubsection*{Free-Flow Speed}
Free-flow speed of the road segment, obtained from speed limit information in the network GIS data. Unit: miles per hour (mph).

\subsubsection*{Number of In-Links}
Number of incoming links to the road segment, derived from the network GIS data.

\subsubsection*{Number of Out-Links}
Number of outgoing links from the road segment, derived from the network GIS data.

\subsubsection*{Orientation}
The orientation of a road link represents the initial compass bearing of the directed link from its upstream node to its downstream node, measured in degrees clockwise from north. This feature is calculated based on the road geometry data.

\subsubsection*{Grade}
The signed percent grade is computed for each directed link as the ratio of elevation change between the downstream and upstream nodes to the horizontal distance. Positive values indicate uphill travel, while negative values indicate downhill travel. Elevation data are obtained from the Digital Elevation Model (DEM) provided by the U.S. Geological Survey (USGS). Unit: percent.

\subsubsection*{Sinuosity}
Sinuosity is defined as the ratio of the polyline length of a road segment to the straight-line distance between its endpoints. This feature is calculated based on the road geometry data.

\subsubsection*{Maximum Curvature}
Maximum curvature is computed by fitting a circle through every three consecutive vertices along the link and taking the inverse of the smallest radius encountered. Links with fewer than three vertices are treated as straight segments and assigned a curvature value of zero. Unit: $1/\text{meter}$. This feature is calculated based on the road geometry data.

\subsubsection*{Mean Curvature}
Mean curvature is computed by fitting a circle through every three consecutive vertices along the link and averaging the inverse of the radius values encountered. Links with fewer than three vertices are treated as straight segments and assigned a curvature value of zero. Unit: $1/\text{meter}$. This feature is calculated based on the road geometry data.

\subsubsection*{Land Use}
Land-use patterns are identified based on parcel-level data. The parcel dataset provides detailed and geographically accurate information on individual land plots, including their boundaries (polygons) and land-use classifications.

For each parcel, the nearest road link(s) are identified based on the minimum Euclidean distance between the parcel geometry and road link geometries. In cases where multiple links have the same minimum distance, all such links are considered nearest links.

For each road link, the set of parcels associated with it is then assembled. The dominant land-use type of the link is determined by aggregating the assigned parcels and selecting the category with the largest total parcel area. This area-weighted approach ensures that larger parcels exert proportionally greater influence on the land-use characterization.

The land-use categories considered in this study include: Residential; General Office; Commercial and Services; Facilities; Education; Military Installations; Industrial; Transportation, Communications, and Utilities; Mixed Commercial and Industrial; Mixed Residential and Commercial; Open Space and Recreation; Agriculture; Vacant; Water; and Unknown.

\subsubsection*{Flow / Previous Flow}
Traffic flow is defined as the number of vehicles passing a road segment at a given location (depending on the detector) during the associated time interval. Unit: vehicles per time interval. Previous Flow: Traffic flow observed during the previous time interval(s).

\subsubsection*{Density / Previous Density}
Traffic density is defined as the number of vehicles occupying a unit length of roadway during the associated time interval. Unit: vehicles per mile per lane. Previous Density: Traffic density observed during the previous time interval(s). 

Density is computed using a commonly adopted approach in transportation research, where density is estimated as the occupancy divided by the average effective vehicle length (i.e., the sum of average vehicle length and detector length). In this study, the average car length is assumed to be 18 ft, the average truck length is 55 ft, and the detector length is 6 ft. The average vehicle length is calculated using the truck proportion as a weighting factor. Occupancy data are obtained from the PeMS dataset.

\subsubsection*{Speed / Previous Speed}
Average traffic speed. Unit: miles per hour (mph).

\subsubsection*{Truck Proportion}
The proportion of trucks in the traffic stream during the associated time interval. Unit: percent.

\subsubsection*{Time of Day}
Time of day represented as a string variable. The temporal resolution in this study is 5 minutes, corresponding to the most granular resolution available in the PeMS dataset.

\subsubsection*{Day of Week}
Day of the week represented as a string variable, with possible values: Mon, Tue, Wed, Thu, Fri, Sat, and Sun.

\subsubsection*{Holiday}
Boolean variable indicating whether the observation corresponds to a public holiday.

\subsubsection*{Weather Type}
Weather condition represented as a categorical string variable, following the coding scheme defined by the National Oceanic and Atmospheric Administration (NOAA).

\subsubsection*{Temperature}
Ambient temperature represented as a continuous variable. Unit: degrees Fahrenheit ($^\circ$F).

\subsubsection*{Visibility}
Visibility represented as a continuous variable, as defined by NOAA. Higher values indicate clearer visibility conditions.

\subsubsection*{Precipitation}
Hourly precipitation represented as a continuous variable, as defined by NOAA. Higher values indicate heavier precipitation.

\par\vspace{0.5em}
The travel time of a road link at a given time interval, used as the prediction target in the fine-tuning phase, is calculated using the length of the road divided by the corresponding traffic speed at that time interval.

\subsection{Matching PeMS Data to Road Links}
PeMS data provide traffic flow, occupancy, and truck proportion at 5-minute intervals, collected at fixed measurement stations. These stations are first matched to road links based on spatial proximity and direction consistency. 

Specifically, for each station, candidate road links are identified within a predefined search radius. Among these candidates, the link with the closest distance and consistent travel direction is selected. In cases where multiple stations are matched to a single road link, the corresponding measurements are aggregated by taking their average values.

\subsection{Matching Speed Data to Road Links}
Speed data are collected for Traffic Message Channel (TMC) segments defined by INRIX. These segments are matched to road links (as defined in the network GIS data provided by the government agency) based on both geometric proximity and directional alignment.

For each TMC segment, candidate road links are first identified within a specified search radius. These candidates are then filtered based on bearing alignment, using an angular threshold to ensure directional consistency. Among the remaining candidates, the road link with the minimum Hausdorff distance to the TMC segment is selected as the match. 

When multiple TMC segments are matched to a single road link, their corresponding speed values are aggregated using the harmonic mean to better account for the characteristics of speed measurements.

\subsection{Weather Data Processing}
Weather data are obtained from NOAA, which provides observations at airport locations. Each road link is matched to its nearest airport, and the corresponding weather data are assigned to the link. The weather-related features used in this study include Dry Bulb Temperature, Hourly Precipitation, Weather Type, and Visibility.

\section{Scenario Settings for Illustration of Dynamic Attention}\label{app:attention-scenarios}

Table \ref{table:sce} lists the input data specifications for the four scenarios used to illustrate the dynamic attention mechanism.

\begin{table}[!htbp]\centering
\caption{Data samples used to illustrate dynamic attention under varying traffic scenarios: (1) a baseline freeway context with typical congestion driven by high traffic flow; (2) an uncongested urban local road in a school zone; (3) an uncongested freeway with an uphill grade and a high truck proportion; and (4) an uncongested freeway with higher horizontal sinuosity.} \label{table:sce}

\scriptsize
\setlength{\tabcolsep}{3pt}
\begin{tabular}{@{}lrrrr@{}}
\toprule
Name & Base Case (Typical Congestion) & School Zone & Truck-Heavy Uphill & High Sinuosity\\
\midrule
Length & 0.5 & 0.5 & 0.5 & 0.5\\
Number of Lanes & 4 & 2 & 4 & 4\\
Free-Flow Speed & 70 & 25 & 70 & 60\\
Number of In-Links & 1 & 3 & 1 & 1\\
Number of Out-Links & 1 & 3 & 1 & 1\\
Orientation & 180 & 180 & 180 & 180\\
Grade & 0 & 0 & 7 & 0\\
Sinuosity & 1 & 1 & 1 & 1.4\\
Maximum Curvature & 0 & 0 & 0 & 0.2\\
Mean Curvature & 0 & 0 & 0 & 0.005\\
Land Use & unknown & Education & unknown & unknown\\
Flow & 820 & 200 & 250 & 400\\
Previous Flow & 810 & 200 & 250 & 400\\
Density & 60 & 30 & 10 & 20\\
Previous Density & 58 & 30 & 15 & 20\\
Previous Speed & 40 & 30 & 65 & 60\\
Truck Proportion & 0 & 0 & 70 & 0\\
Time of Day & 09:00 & 09:00 & 09:00 & 09:00\\
Day of Week & Wed & Wed & Wed & Wed\\
Holiday & False & False & False & False\\
Weather Type & clear & clear & clear & clear\\
Temperature & 70 & 70 & 70 & 70\\
Visibility & 10 & 10 & 10 & 10\\
Precipitation & 0 & 0 & 0 & 0\\
\bottomrule
\end{tabular}
\end{table}

\FloatBarrier

\end{appendices}

\backmatter

\bmhead{Code and data availability}

All code will be shared in a public repository. All data will be shared, except for the raw data from private-sector providers such as INRIX.

\bmhead{Acknowledgements}

This research is in part supported by U.S. National Science Foundation CMMI-2528901, and Safety21, a National University Transportation Center on Safety sponsored by U.S. Department of Transportation.

\bmhead{Author contributions}

X.M. and S.Q. designed research; X.M. collected and processed data and performed research; X.M. wrote the draft; and S.Q. reviewed and edited the paper.

\bmhead{Competing interests}

The authors declare no competing interests.

\bibliography{references}

\end{document}